\documentclass[superscriptaddress,twocolumn,10pt]{revtex4-1}

\usepackage{graphicx,color,amsmath}

\usepackage{listings} 

\lstnewenvironment{alg}[1][]
{\lstset{language={[95]Fortran},frame=single,mathescape=true,commentstyle=\normalfont,basicstyle={\ttfamily},captionpos=b,linewidth=0.9\columnwidth,xleftmargin=12pt,xrightmargin=-14pt,framesep=11pt,escapechar=\%,float=h!,#1 }}
{}

\begin{document}

\title{Constant-pressure nested sampling with atomistic dynamics} 

\author{Robert J.N. Baldock} 
\email{rjnbaldock@gmail.com}
\affiliation{Theory and Simulation of Materials (THEOS), and National Centre for Computational Design and Discovery of Novel Materials (MARVEL), \'Ecole Polytechnique F\'ed\'erale de Lausanne, CH-1015 Lausanne, Switzerland}
\author{Noam Bernstein} 
\affiliation{Center for Materials Physics and Technology, Naval Research Laboratory, Washington, DC 20375, USA}
\author{K. Michael Salerno}
\affiliation{National Research Council Associateship Program, resident at the U.S. Naval Research Laboratory, Washington DC 20375, USA}
\author{L\'{\i}via B. P\'artay} 
\affiliation{Department of Chemistry, University of Reading, United Kingdom}
\author{G\'abor Cs\'anyi} 
\affiliation{Engineering Laboratory, University of Cambridge, United Kingdom}

\newcommand*\nb[1]{{\color{blue}\tt#1}}
\newcommand*\csg[1]{{\color{green}#1}}
\newcommand*\CSG[1]{{\color{green}\tt#1}}
\newcommand*\lbp[1]{{\color{red}\tt#1}}
\newcommand*\rjnb[1]{{\color{cyan}\tt#1}}

\begin{abstract}

The nested sampling algorithm  has been shown to
be a general method for calculating the pressure-temperature-composition
phase diagrams of materials.  While the previous implementation
used single-particle Monte Carlo moves,
these are inefficient for condensed systems with general interactions
where single-particle moves cannot be evaluated faster than the energy of the whole system.
Here we enhance the method by using  all-particle moves: either Galilean Monte Carlo or
a total enthalpy Hamiltonian Monte Carlo algorithm, introduced in this paper.
We show that these algorithms
enable the determination of phase transition temperatures with equivalent accuracy
to the previous method at $1/N$ of the cost for an $N$-particle
system with general interactions, or at equal cost when single
particle moves can be done in $1/N$ of the cost of a full $N$-particle
energy evaluation.
We demonstrate this speedup for the freezing and condensation transitions of the Lennard-Jones system and show the utility of the algorithms by calculating the order-disorder phase transition of a binary Lennard-Jones model alloy, the eutectic of copper-gold, the density anomaly of water and the condensation and solidification of bead-spring polymers.
The nested sampling method with all three algorithms is implemented in the \texttt{pymatnest} software.

\end{abstract}

\maketitle

\section{Introduction}

The ability to predict the behavior of materials under a variety of conditions is important in both academic and industrial settings.
In principle, statistical mechanics enables the prediction of the properties of materials in thermodynamic equilibrium from the microscopic interaction of atoms.
Computer simulation is the tool for using numerical statistical mechanics in practice, and a wide variety of models and approximations are used for atomic interactions, all the way from the electronic Schr\"odinger equation to simple hard spheres.
The equilibrium pressure-temperature phase diagram for a given composition is one of the most fundamental properties of a material, and forms the basis for making, changing, designing, or in general, just thinking about the material.  

The common  approach is to use different computational methods for resolving each of the transitions between the phases, or for comparing the stability of particular combinations of phases. 
This requires the prior knowledge of a  list of proposed phases or crystal structures  at each set of thermodynamic parameters.
In a previous paper~\cite{pt_phase_dias_ns} we introduced a nested sampling (NS) algorithm~\cite{bib:skilling,bib:skilling2} that enables the automated calculation of complete pressure-temperature-composition phase diagrams.
The nested sampling algorithm constitutes a single method for resolving all phase transitions automatically in the sense that no prior knowledge of the phases is required.

The algorithm in~\cite{pt_phase_dias_ns} used Gibbs sampling, i.e. single particle Monte Carlo (SP-MC) moves, to explore configuration space. 
For some interactions, which we term {\em separable}, the energy change of an $N$-particle system due to the displacement of a single 
particle can be calculated in $1/N$ times the cost of evaluating the energy of the whole system. 
Thus, for separable interactions, the cost of an $N$-particle sweep is equal to the cost of a single full-system energy/force evaluation.
For these cases Gibbs sampling is efficient.
However, interactions in general are not separable, and if a single-particle move is just as costly as a full-system evaluation, a sweep that moves every particle costs $N$ times more than a single full-system energy/force calculation.  
All-particle MC moves can be used
in principle, but it is well known that making such moves in random directions leads to very slow exploration
in condensed phases (liquids, solids), because maintaining a finite MC acceptance rate requires that the displacement of
each atom become smaller as the system size increases~\cite{FrenkelBookManyParticleMC}.

Here we replace Gibbs sampling by either one of two algorithms that use efficient all-particle moves inspired by the Hamiltonian Monte Carlo (HMC) method~\cite{Duane_HMC_87}. The purpose of this paper is to show how these algorithms can be utilized efficiently in the case of nested sampling. One is a total enthalpy HMC (TE-HMC) introduced in this paper, which takes advantage of standard molecular dynamics (MD) implemented in many simulation packages.
The other is Galilean Monte Carlo (GMC)~\cite{Skilling_GMC, betancourt_ns_chmc}, which is not 
as widely available, but does not suffer from the breakdown of ergodicity that the TE-HMC algorithm may
experience for systems with a large number of particles. We compare the efficiency of these two methods with that of Gibbs sampling, and find 
that they require comparable numbers of whole-system energy/force evaluations, leading to a factor of $1/N$
reduction in computational time for general inter-particle interactions.

The GMC and TE-HMC algorithms use inter-particle force information to move all particles
coherently
along ``soft'' degrees of freedom, and therefore 
explore faster than simple diffusion, at least over short-time scales. 
Effectively, short trajectories are all-particle MC move proposals with large
step lengths that still lead to reasonable MC acceptance probabilities. 
Over the time scale of an entire Markov chain Monte Carlo (MCMC)
walk,
the
motion is still diffusive, but the short time coherence helps GMC
and TE-HMC explore configuration space much faster than randomly oriented
all-particle moves.
Using GMC or TE-HMC enables the simulation of  a wide range of phase transitions in atomistic and particle systems with NS, including chemical ordering in binary Lennard-Jones (LJ), the eutectic of copper-gold alloys,
freezing of water, and the transitions of a coarse-grained bead-spring
polymer model.  Parallel implementations of both algorithms are available in the {\tt pymatnest}  python software
package~\cite{pymatnest}, using the LAMMPS package~\cite{LAMMPS} for the dynamics.

\section{The nested sampling method} 
\label{sec:method}

The nested sampling algorithm with constant pressure and flexible boundary conditions (i.e. with variable periodic cell shape and volume) calculates  the cumulative density of states, $\chi(\widetilde{H}),$ at fixed pressure $P$, where $\widetilde{H}=U+PV$ is the configurational enthalpy, $U(\mathbf{r})$ is the potential energy function, and $V$ is the volume of the system.
From the cumulative density of states, one can calculate the partition function and heat capacity as explicit functions of temperature. 
Nested sampling also returns a series of atomic configurations, from which one may compute ensemble averages of observables and free energy landscapes~\cite{pt_phase_dias_ns}.

The key idea of nested sampling is that it constructs a series of decreasing enthalpy levels $\{\widetilde{H}^\mathrm{sup}_i\}$, each of which bounds from above a volume of configuration space $\chi_i$, with the property that $\chi_i$ is approximately a constant factor smaller than the volume $\chi_{i-1}$ of the level above.
The constant pressure partition function, and its approximation in nested sampling, are then given by
\begin{align}
\Delta\left(  N, P, \beta  \right) &= \frac{\beta P}{N!}\left(\frac{2\pi m}{\beta h^2} \right)^{3N/2}\int_{-\infty}^\infty{d\widetilde{H} \frac{\partial \chi}{\partial \widetilde{H}} e^{-\beta \widetilde{H}} } \label{eq:npt_pf} \\
& \approx \frac{\beta P}{N!}\left(\frac{2\pi m}{\beta h^2} \right)^{3N/2}\sum_i \left( \chi_{i-1} -\chi_i \right)e^{-\beta \widetilde{H}^\mathrm{sup}_i} \label{eq:ns_approx_npt_pf} .
\end{align}
Here $N$ is the number of particles of mass $m$, $h$ is Planck's constant and  $\partial \chi/\partial \widetilde{H}$  is the density of enthalpy states.

To calculate the absolute value of the partition function~\eqref{eq:ns_approx_npt_pf}, we must possess the absolute values of the configuration space volumes $\{\chi_i\}$.
The volumes $\{\chi_i\}$ are specified in NS as a decreasing geometric progression, starting from $\chi_0$, which is the total (finite) volume of the configuration space.
The configuration space must therefore be compact.
In order to ensure that we are sampling from a compact configuration space, the simulation cell volume $V$ is restricted to be less than a maximum value $V_0$, chosen to be sufficiently large as to correspond to an almost ideal gas.
The total configuration space volume is therefore $\chi_0=V_0^{N+1}/(N+1)$.
Restricting the sampling to $V<V_0$  allows a good approximation of the partition function 
provided $k_\mathrm{B}T\ll P V_0$~\cite{pt_phase_dias_ns}.

The simulation cell is periodic, and is represented by $\mathbf{h}$, a $3\times3$ matrix of lattice vectors that relates the Cartesian positions of the particles $\mathbf{r}$ to the fractional coordinates $\mathbf{s}$ by $\mathbf{r}=\mathbf{hs}$.
The volume of the simulation cell is $V=\det\mathbf{h}$, and $\mathbf{h}_0=\mathbf{h}V^{-1/3}$ is the image of the unit cell normalized to unit volume.  The NS algorithm maintains a pool of $K$ configurations drawn from 
\begin{equation}  \label{eq:fixedP_ns_dist}
\begin{aligned} 
&\mathrm{Prob}\left( \mathbf{s}, \mathbf{h}_0, V | V_0, \widetilde{H}^\mathrm{sup}, d_0   \right)  \propto V^N \delta\left(\det\mathbf{h}_0 -1\right)  \\ 
& \times  \Theta\left(\widetilde{H}^\mathrm{sup} - \widetilde{H} \right) \Theta\left(V_0-V\right) \\
& \times \prod_{i=1}^{3N}{\Theta(s_i)\Theta(1-s_i)} \, \Theta\left( d_0-\min_{j \neq k}\left[ \frac{1}{\mathbf{h}_0^j \times \mathbf{h}_0^k}  \right]\right) ,
\end{aligned}
\end{equation}
where $\Theta$ is the Heaviside step function, $\mathbf{h}_0^i$ are columns of $\mathbf{h}_0$, and $\widetilde{H}^\mathrm{sup}$ is a maximum configurational
enthalpy.  Thus, fractional particle coordinates $\mathbf{s}$  are uniformly distributed on $(0,1)^{3N}$, $\widetilde{H}$ is restricted to be
smaller than $\widetilde{H}^\mathrm{sup}$, and $V$ is restricted to be smaller than  $V_0$.
The last term in Eq.~\eqref{eq:fixedP_ns_dist} restricts
the simulation cell from becoming too thin (controlled by the parameter $d_0$), thus avoiding unphysical
correlations between interacting periodic images~\cite{pt_phase_dias_ns}.
For simple fluids of 64 atoms, $d_0$ should be set no smaller than 0.65, while simulations with larger numbers of atoms can tolerate smaller values of $d_0$~\cite{pt_phase_dias_ns} (see Appendix~\ref{app:settingd0}).
The probability distribution~\eqref{eq:fixedP_ns_dist} corresponds to a uniform distribution over the Cartesian particle coordinates $\mathbf{r}$, subject to the constraints above.

The simulation is initialized by drawing $K$ configurations from distribution~\eqref{eq:fixedP_ns_dist}, with $\widetilde{H}^\mathrm{sup}=\infty$.
After initialization, the NS algorithm performs the following loop, starting with $i=1$.
\begin{enumerate}

    \item \label{nsalg:recordlevels} From the set of $K$ configurations,
    record the $K_r \ge 1$ samples with the highest configurational
    enthalpy, $\{\widetilde{H}\}_i$. 
    Use the lowest enthalpy from that set,
    $\min \{\widetilde{H}\}_i$, as the new enthalpy limit:
    $\widetilde{H}^\mathrm{sup}\leftarrow \min\{\widetilde{H}\}_i$.
    The volume of configuration space with enthalpy equal to or less than
    $\widetilde{H}^\mathrm{sup}$ is $\chi_i\approx \chi_0[(K-K_r+1)/(K+1)]^i$.

    \item \label{nsalg:removeconfigs} Remove the $K_r$ samples with enthalpies $\{\widetilde{H}\}_i$
    from the pool of samples and generate $K_r$ new configurations from the
    distribution~\eqref{eq:fixedP_ns_dist}, using the updated value of $\widetilde{H}^\mathrm{sup}$.  
    This is achieved by first
    choosing $K_r$ random configurations from the pool of remaining samples,
    creating clones of those configurations, and evolving the cloned
    configurations using a MCMC algorithm that converges
    to the distribution~\eqref{eq:fixedP_ns_dist}.

    \item Set $i\leftarrow i+1$, and return to step 1 unless a stopping criterion is 
       met (see Appendix~\ref{app:Testimate}).
\end{enumerate}

The required values of $K$ and the MCMC walk-length, $L$, depend on the system being studied.
This behavior is described in Refs.~\cite{bib:skilling,bib:skilling2,bib:our_NS_paper,pt_phase_dias_ns}.
In addition, increasing $K_r$ allows for greater parallelization of the NS algorithm; however, increasing $\frac{K_r-1}{K}$ also leads to greater error in the estimates of $\{\chi_i\}$, so care should be taken not to make $\frac{K_r-1}{K}$ too large (see Appendix~\ref{app:para}).

Thermodynamic expectation values and free energy landscapes can be computed using the samples recorded during step~\ref{nsalg:recordlevels} of the NS algorithm~\cite{pt_phase_dias_ns}.
Representative configurations can be sampled at any temperature simply by choosing configurations at random, according to their thermalized probabilities
\begin{equation}
p_i\left(\beta\right) = \frac{ \left(\chi_{i-1} - \chi_{i} \right) e^{-\beta \widetilde{H}_i} }{\sum_i{\left(\chi_{i-1} - \chi_{i} \right) e^{-\beta \widetilde{H}_i}} }.
\end{equation}
Examining a small number of configurations chosen in this way is often sufficient to understand which phase occurs at each temperature.
This method was used to choose the atomic configurations shown in Sec.~\ref{sec:results}.

\section{Markov chain Monte Carlo algorithms \label{sec:MCMCsteps}}

To decorrelate the cloned configurations in step~\ref{nsalg:removeconfigs} of the NS algorithm we use a MCMC algorithm that converges to the  distribution~\eqref{eq:fixedP_ns_dist}
by applying two kinds of steps: cell steps, including changes to volume
and shape, and particle steps, including continuous motion in space and
(for multicomponent systems) coordinate swaps between particles of different types.  

The cell steps
include volume steps that ensure 
$\mathrm{Prob}(V)\propto V^N$, 
and shearing and stretching steps that lead to
$\mathrm{Prob}(\mathbf{h}_0)\propto \delta\left(\det\mathbf{h}_0 -1\right) \Theta\left( d_0-\min_{i \neq j}\left[ \frac{1}{\mathbf{h}_0^i \times \mathbf{h}_0^j}  \right]\right)$, 
as required by the target distribution~\cite{pt_phase_dias_ns}. The following subsections introduce the algorithms for moving the configuration in the space of the atomic coordinates. 

\subsection{Galilean Monte Carlo \label{sec:GMC}}
\label{sec:MCMCsteps:GMC}

In GMC~\cite{Skilling_GMC, betancourt_ns_chmc} one defines an infinite square-well potential function $\widetilde{H}_\mathbf{GMC}$,
\begin{equation} \label{eq:GMC_PE}
\begin{split}
\widetilde{H}_\mathbf{GMC}\left(\mathbf{s},V,\mathbf{h}_0\right) =& 
\begin{cases} 
0\: &:\:\widetilde{H} < \widetilde{H}^\mathrm{sup}, \\
\infty   \:   &:\:\widetilde{H} \ge \widetilde{H}^\mathrm{sup} , \\
\end{cases}
\end{split}
\end{equation} 
which is equal to the logarithm of the desired probability distribution: uniform over the allowed region, $\widetilde{H} < \widetilde{H}^\mathrm{sup}$, and zero elsewhere.
Note that we have omitted the constraints on $V$ and $\mathbf{h}_0$ from~\eqref{eq:GMC_PE} since we use GMC to explore only the atomic coordinates.
Having defined $\widetilde{H}_\mathbf{GMC}$, one samples the fractional atomic coordinates $\mathbf{s}$ uniformly by performing standard Hamiltonian Monte Carlo sampling~\cite{Duane_HMC_87} on the function $\widetilde{H}_\mathbf{GMC}$.
We follow the GMC approach proposed by Betancourt~\cite{betancourt_ns_chmc}, which uses a fixed number of force evaluations and therefore helps the load balance when parallelizing the algorithm (see Appendix~\ref{app:para}).
Our implementation, expressed for practical convenience of implementation in Cartesian coordinates $\mathbf{r}$ rather than fractional coordinates $\mathbf{s}$, is as follows.

At the start of each atomic GMC trajectory, save the initial atomic coordinates $\mathbf{r}_0$.
Generate a velocity $\mathbf{v}$ chosen uniformly from the surface of a $3N$-dimensional hypersphere of radius 1. 
Repeat the following loop $L$ times:
\begin{enumerate}
\item Propagate the atomic coordinates $\mathbf{r}$ in the direction of $\mathbf{v}$ for one step of length $dt$. 
The atomic coordinates are now $\mathbf{r}^*$.
   \item If $\widetilde{H} \ge \widetilde{H}^\mathrm{sup}$, attempt to redirect  the trajectory 
      back into the allowed region by reflecting velocities from the current position, by
      $\mathbf{v} \leftarrow \mathbf{v} - 2(\mathbf{v}\cdot\hat{\mathbf{n}}) \hat{\mathbf{n}}$,
      where $\hat{\mathbf{n}}=-\nabla_{\mathbf{r}}{\widetilde{H}}/|\nabla_{\mathbf{r}}{\widetilde{H}}|$.
Following velocity reflection,  propagation continues from $\mathbf{r}^*$.
\end{enumerate}
Finally, if at the end of the trajectory $\widetilde{H} \ge \widetilde{H}^\mathrm{sup}$, reject the trajectory and return to the initial atomic coordinates $\mathbf{r}_0$. Note that the reflection in step 2 {\em does not} occur at the exact boundary $\widetilde{H} = \widetilde{H}^\mathrm{sup}$, and this is essential to maintain detailed balance. 
The acceptance rate of the GMC step is controlled by adjusting $dt$. There are similarities between GMC and the ``hit and run'' algorithm for sampling convex volumes~\cite{hitandrunSmith}, and it remains to be seen whether the clear advantages over Gibbs sampling have common underlying reasons~\cite{Lovasz1999}.  

\subsection{Total enthalpy Hamiltonian Monte Carlo \label{sec:TE-HMC}}

One disadvantage of GMC is that
when the boundary of the allowed region of configuration space $\widetilde{H}<\widetilde{H}^\mathrm{sup}$ is
 complicated, attempts to reflect the sampler back into the allowed region fail frequently,
and reflection
further into the disallowed region often leads to 
rejection of the entire trajectory, thus overall driving down the optimal step size.  
Hamiltonian (constant energy) molecular dynamics, on the other hand, can
generate nearly constant {\em total} energy trajectories using comparatively large step sizes, and use the
exchange of energy between potential and kinetic degrees of freedom
to smooth the ``reflections'' from high potential energy regions.  
In TE-HMC we take advantage of this behavior
by using short MD
trajectories to evolve the atomic coordinates.

Hamiltonian dynamics couples the evolution of the atomic momenta and coordinates, and
in TE-HMC we explicitly sample the {\em total  phase space}  of the atoms $(\mathbf{s},V,     \mathbf{h}_0,\mathbf{p})$, where $\mathbf{p}$ denotes the Cartesian momenta.
In contrast, in SP-MC and GMC one samples only the atomic {\em configuration} space $(\mathbf{s},V,\mathbf{h}_0)$. 

In step 1 of each NS iteration the $K_r$ samples with highest total enthalpy $H=(\widetilde{H}+E_k)$, where $E_k(\mathbf{p})$ is the kinetic energy, are identified as the next set of recorded samples. 
Next, the total enthalpy limit is updated $H^{\mathrm{sup}} \leftarrow \min\{H\}_i$, and in step 2 $K_r$ new samples are generated from the joint probability distribution 
\begin{equation}  \label{eq:fixedP_ns_dist_tehmc} 
\begin{aligned} 
&\mathrm{Prob}\left( \mathbf{s}, \mathbf{h}_0, V, \mathbf{p} | V_0, H^\mathrm{sup} , d_0, E_k^0  \right) \propto  V^N \delta\left(\det\mathbf{h}_0 -1\right) \\
& \times \Theta\left(H^\mathrm{sup} - H \right) \Theta\left(V_0-V\right) \Theta\left(E_k^0 - E_k\left(\mathbf{p}\right) \right) \\
&\times  \prod_{i=1}^{3N}{\Theta(s_i)\Theta(1-s_i)} \Theta\left( d_0-\min_{j \neq k}\left[ \frac{1}{\mathbf{h}_0^j \times \mathbf{h}_0^k}  \right]\right) .
\end{aligned}
\end{equation}
Distribution~\eqref{eq:fixedP_ns_dist_tehmc} invokes
the same constraints on $V$, $\mathbf{h}_0$ as  distribution~\eqref{eq:fixedP_ns_dist}, but restricts $H<H^\mathrm{sup}$ and specifies that the 
momenta are uniformly distributed in the region
\begin{equation}\label{eq:ke_maxval}
E_k(\mathbf{p})<E_k^0.
\end{equation}
Using both maximum volume and kinetic energy values $V_0$ and $E_k^0$ ensures that the phase space we sample is compact, the necessity of which is explained in Sec.~\ref{sec:method}.
In particular, $V_0$ and $E_k^0$ enforce compactness of the sampled configuration and momentum spaces, respectively.

We initialize exactly as described in Sec.~\ref{sec:method}, except that we now assign each sample momenta chosen uniformly at random from the region~\eqref{eq:ke_maxval}.
This is achieved using Algorithm~\ref{alg:ns_tot_p_rand} given in Appendix~\ref{app:te_hmc}.
We choose $E_k^0=\frac{3}{2}PV_0$ as for an ideal gas, so that, again, we obtain a good approximation of the partition function provided $k_{\mathrm{B}}T \ll P V_0$.
Probability distribution~\eqref{eq:fixedP_ns_dist_tehmc} corresponds to a uniform distribution over the phase space coordinates of the system $(\mathbf{s},\mathbf{p})$, subject to the above constraints.
[Recall that, in contrast,  probability distribution~\eqref{eq:fixedP_ns_dist} 
in Sec.~\ref{sec:method} corresponds to a uniform distribution over the particle coordinates $\mathbf{s}$ alone, subject to similar constraints.]

Since in TE-HMC we explicitly sample both coordinates and momenta, the nested sampling approximation to the partition function becomes
\begin{equation}
\Delta\left(  N, P, \beta  \right)  \approx \frac{\beta P}{N!h^{3N}} \sum_i \left( \Gamma_{i-1} -\Gamma_i \right)e^{-\beta H_i}, \label{eq:hmc_ns_approx_npt_pf}
\end{equation}
where $H_i$ is the total enthalpy of the $i^\mathrm{th}$ nested sampling level, and $\Gamma_i$ 
is the volume of phase space with total enthalpy less than or equal to $H_i$ at pressure $P$.

In order to ensure the sampler spends approximately an equal amount of computer time exploring each degree of freedom, we set all the masses to be equal: $m_i=m \: \forall \: i$.
We then recover the correct partition function~\eqref{eq:hmc_ns_approx_npt_pf} by multiplication
\begin{equation}
\Delta\left(  N, P, \beta  \right) \approx  \left(\prod_{i=1}^{N}{\frac{m_i}{m}}\right)^{\frac{3}{2}}\Delta_{\mathrm{NS}}\left(  N, P, \beta  \right), 
\end{equation}
where $\Delta_{\mathrm{NS}}$ is equal to the right hand side of~\eqref{eq:hmc_ns_approx_npt_pf}, calculated with equal particle masses.
For equal particle masses, we find $\Gamma_0$ in Eq.~\eqref{eq:hmc_ns_approx_npt_pf} to be given by  
\begin{equation}\label{eq:npt_gamma0}
\Gamma_0 = \frac{V_0^{N+1}}{N+1} \frac{2 \left(2 \pi m E_k^0  \right)^{\frac{3N}{2}} }{3 N \Gamma\left( \frac{3N}{2} \right) }
\end{equation}
where $\Gamma\left( \frac{3N}{2} \right)$ is the gamma function evaluated at $\frac{3N}{2}$.

In TE-HMC, the atomic coordinates $\mathbf{r}$ and momenta $\mathbf{p}$ are evolved according to the following Hamiltonian Monte Carlo sequence.
The move begins with the initial Cartesian phase space coordinates $(\mathbf{r}^{(0)},\mathbf{p}^{(0)})$, which are in the allowed region, $H<H^{\mathrm{sup}}$.
\begin{enumerate}
\setlength\itemsep{-0.3em}
\item Randomize the momenta, either partially or completely, to pick new momenta satisfying $E_k(\mathbf{p})<\min[E_k^0,H^\mathrm{sup}-\widetilde{H}]$, as in Eq.~\eqref{eq:fixedP_ns_dist_tehmc}.
This momentum randomization takes us to the coordinates $(\mathbf{r}^{(0)},\mathbf{p}^{(1)})$.
\item \label{enumerate:newtonstep} Starting from $(\mathbf{r}^{(0)},\mathbf{p}^{(1)})$, integrate Newton's equations of motion for the coordinates $(\mathbf{r},\mathbf{p})$ over a fixed number of time steps.
At the end of this trajectory the phase space coordinates are $(\mathbf{r}^{(1)},\mathbf{p}^{(2)})$.
\item Reverse the momenta $\mathbf{p}^{(3)}=-\mathbf{p}^{(2)}$;
the trajectory and this reversal taken together, $(\mathbf{r}^{(0)},\mathbf{p}^{(1)})\rightarrow (\mathbf{r}^{(1)},\mathbf{p}^{(3)})$, are a reversible MC proposal, which ensures that the move satisfies detailed balance.
\item Calculate the new total enthalpy $H_\mathrm{trial}=H(\mathbf{r}^{(1)},V,\mathbf{h}_0,\mathbf{p}^{(3)})$.
If $H_\mathrm{trial}<H^\mathrm{sup}$ and $E_k(\mathbf{p}^{(3)})<E_k^0$ then accept the new coordinates $(\mathbf{r}^{(1)},\mathbf{p}^{(3)})$, otherwise return to the starting coordinates $(\mathbf{r}^{(0)},\mathbf{p}^{(0)})$.
The coordinates are now $(\mathbf{r}^*,\mathbf{p}^*)$.
\item Reverse the momenta again.
The final, resulting coordinates are $(\mathbf{r}^*,-\mathbf{p}^*)$.
\end{enumerate}

A great advantage of the TE-HMC algorithm is that numerical integration of Newton's equations of motion approximately conserves the total enthalpy along the trajectory such that the value only fluctuates by a small amount.
Consequently, the trial coordinates $(\mathbf{r}^{(1)},\mathbf{p}^{(3)})$ nearly always satisfy $H_\mathrm{trial}<H^\mathrm{sup}$.
Thus if partial momentum randomization is used, better preserving the direction of motion of the particles, one obtains excellent continuation between successive short TE-HMC trajectories. Pseudocode for the TE-HMC move introduced above is given in Appendix~\ref{app:te_hmc}. 
In particular, Algorithm~\ref{alg:ns_partial_p_rand} introduces the parameter $\gamma$, which controls the extent to which the direction of motion of the particles is randomized when using partial momentum randomization.

\subsubsection{Wider application of NS with TE-HMC}

The TE-HMC method can be used in applications of NS as a general method for Bayesian computation~\cite{bib:skilling,bib:skilling2}, by setting $ m=1 $, equating $\widetilde{H}(\mathbf{r}) = - \ln{\mathrm{Prob}(\mathbf{r})}$ where $\mathrm{Prob}(\mathbf{r})$ is the Bayesian likelihood of the parameters $\mathbf{r}$, and specifying a suitably large value for the maximum kinetic energy, $E_k^0$.
The Bayesian evidence is approximated by
\begin{equation}
Z \approx \sum_i \left( \Gamma_{i-1} -\Gamma_i \right)e^{-H_i}
\end{equation}
in which $\Gamma_0$ is given by
\begin{equation}
\Gamma_0 = \frac{2  {E_k^0} ^{\frac{d}{2}} }{3 N \Gamma\left( \frac{d}{2} \right) }
\end{equation}
where $\Gamma\left( \frac{d}{2} \right)$ is the gamma function evaluated at $\frac{d}{2}$, and $d$ denotes the dimension of the parameter space: the phase space $\{\mathbf{r},\mathbf{p}\} $ has $2d$ dimensions.
Throughout this paragraph we have assumed that the prior over $\mathbf{r}$ is uniform. For continuous problems it is always possible to use coordinates in which this is the case.

A total {\em energy} HMC algorithm, suitable for performing constant volume NS calculations, is obtained by replacing $\beta P$ with $1$ in Eq.~\eqref{eq:hmc_ns_approx_npt_pf} then setting the pressure $P = 0$ throughout the TE-HMC algorithm. 
Cell volume MC moves should not be performed, while cell stretch and cell shear moves may optionally be included or left off. 
Momenta are initialized as for total enthalpy HMC, with $E_k^0 = \frac{3N}{2} k_B T_{0}$ where $T_0$ is a high temperature which corresponds to the ideal gas. 
In total energy HMC one must replace Eq.~\eqref{eq:npt_gamma0} with 
\begin{equation}
\Gamma_0 = V^N \frac{2 \left(2 \pi m E_k^0  \right)^{\frac{3N}{2}} }{3 N \Gamma\left( \frac{3N}{2} \right) }.
\end{equation}
This total energy HMC algorithm was used to perform the constant volume calculations reported in Sec.~\ref{sec:res_mol_solid}.

\section{Results}
\label{sec:results}

\subsection{Parameters and implementation}
\label{sec:results:params}

Here we present tests of the performance of the constant pressure NS sampling
method  with the different particle motion algorithms.
The  single-particle MC moves (SP-MC) are grouped into 
sweeps over the system moving each particle in random order.  
When the potential is separable, for example the LJ 
model in the examples below, the cost of an entire $N$-particle sweep
is equal to a single full-system energy/force evaluation.
When discussing walk lengths below we therefore consider an
$N$-particle sweep equivalent to a single energy/force evaluation in
other moves (GMC or TE-HMC for particle positions, as well as cell moves).
For interactions of more general form that are not separable,
the sweep would be $N$ times slower than an energy/force evaluation.

Step sizes for all
types of MCMC moves are automatically adjusted during the NS iteration
process using pilot walks (which are not included in the NS configuration
evolution) so as to reach acceptance rates of 0.5-0.95 for TE-HMC MD
trajectories, and 0.25-0.75 for all other moves (cell steps, single particle SP-MC
steps, GMC trajectories).  The essential NS parameters for all
systems presented here are listed in Table~\ref{table:params},
and input files are provided in the Supplemental Material (SM)~\cite{SI}.
Each particle step
consisted of 8 $N$-particle energy/force evaluations: 8 all-particle sweeps for
SP-MC, or a single 8 step trajectory for GMC and TE-HMC.  For multicomponent
systems 8 swap steps were done in addition~\cite{pt_phase_dias_ns}.   For TE-HMC partial
randomization of the velocity direction was done as in 
Algorithm~\ref{alg:ns_partial_p_rand} with $\gamma = 0.3$, except for the polymer  system
which used $\gamma = 0.1$.

\begin{table*}
\caption{Parameters for NS runs: pressure $P$, number of particles $N$, number of configurations $K$,
number of configurations removed per iteration $K_r$, walk length (all-particle  energy/force calls) $L$,
step ratios (particle : cell volume : cell shear : cell stretch [: swap]),
minimum temperature $T_\mathrm{min}$, and number of NS iterations $n_\mathrm{iter}$.
Each particle step
consisted of 8 $N$-particle energy/force evaluations: 8 all-particle sweeps for
SP-MC, or a single 8 step trajectory for GMC and TE-HMC.
}
\label{table:params}
\begin{ruledtabular}
\begin{tabular}{lcccccccc}
    & $P$                                       & $N$           & $K$   & $K_r$ & $L$       & step ratios
    & $T_\mathrm{min}$  & $n_\mathrm{iter}/10^6$  \\
mono LJ walk length 
    & $3.162 \times 10^{-2}~\epsilon/\sigma^3$  & 64            & 2304  & 1     & 80-2560    & 1:16:8:8
    & 0.43~$\epsilon$              &  $2 $ \\
binary LJ order-disorder
    & $3.162 \times 10^{-2}~\epsilon/\sigma^3$  & 64            & 4608  & 2     & 1536       & 1:16:8:8:8
    & 0.043~$\epsilon$             & $3 $ \\
Cu, Au
    & 0.1~GPa                                   & 64            & 2304  & 1     & 768        & 1:16:8:8:8
    & 600~K             & $2 $ \\
Cu$_x$Au$_{1-x}$ $x = \left[ 0.25, 0.5, 0.75 \right]$
    & 0.1~GPa            & 64            & 4608  & 2     & 768        & 1:16:8:8:8
    & 600~K                 & $2$ \\
mW water
    & 1.6~MPa                                   & 64            & 1920  & 1    & 3168       & 3:8:4:4
    & 150~K                 & $2$ \\
single chain polymer 
    & const. V                    & 15            & 2304  & 1     & 5120      & 1:0:0:0
    & 0.01~$\epsilon$              & $1$ \\
multichain polymer cluster 
    & const. V                             & $8 \times 15$ & 4608  & 1     & 5120     & 1:0:0:0
    & 0.3~$\epsilon$               & $8$ \\
multichain polymer
    & $2.3 \times 10^{-3}~\epsilon/\sigma^3$    & $8 \times 15$ & 4608  & 1     & 5120     & 1:4:4:4
    & 1.2~$\epsilon$               & $3$
\end{tabular}
\end{ruledtabular}
\end{table*}

\subsection{Efficiency of the SP-MC, GMC, and TE-HMC algorithms}
\label{sec:results:mono_LJ_walk_length}

\begin{figure}
    \centerline{\includegraphics[width=0.9\columnwidth]{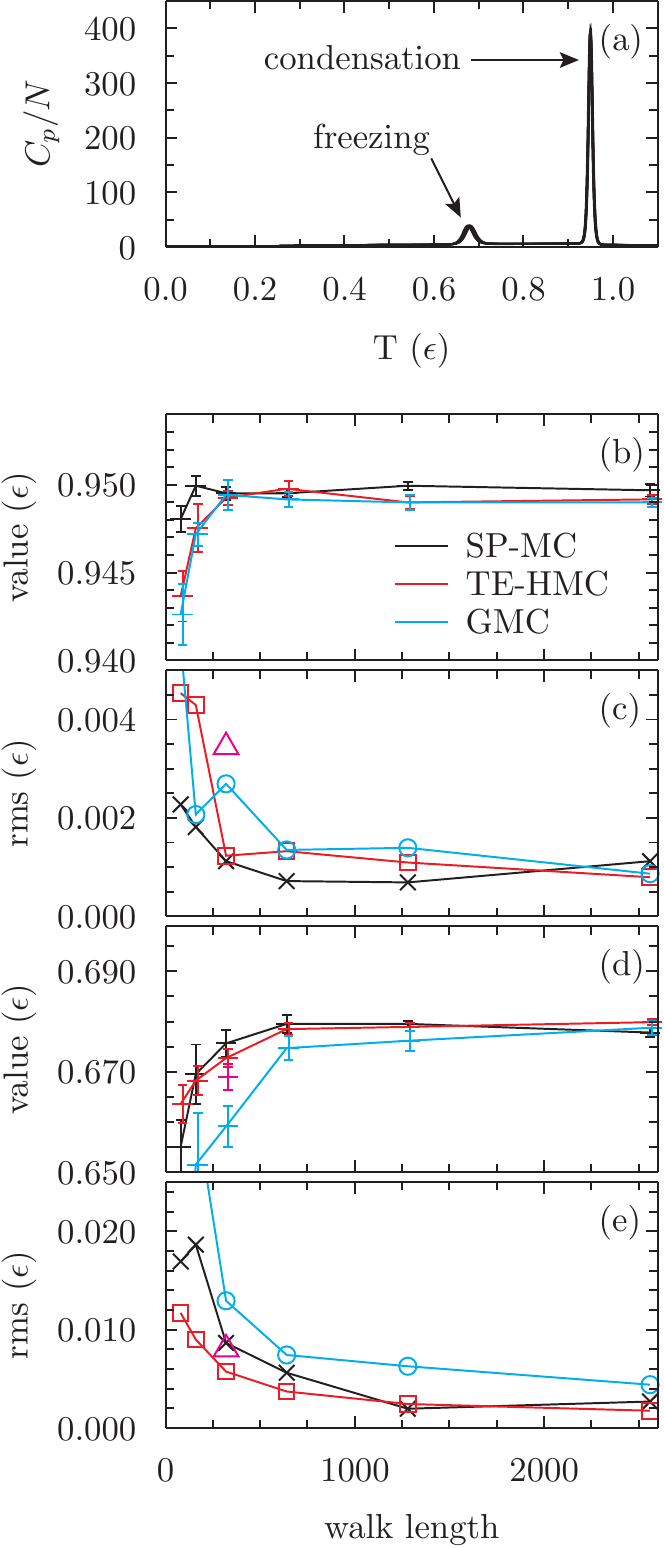}}

    \caption{Heat capacity curves $C_p(T)$, peak positions, and peak scatter
    calculated using NS for 64 LJ particles 
    at pressure $P \sigma^3/\epsilon = 3.162\times 10^{-2}$.
    Panel (a): Example $C_p(T)$ curve for 10 independent runs with SP-MC  moves and $L=
1280$ energy evaluations. Panels (b) -- (e):
    Convergence of the condensation [(b), (c)] and freezing
    [(d), (e)] transition peak temperatures as a function of walk length.
    For each transition, the upper panel [(b), (d)] shows the mean peak
    position (error bars indicate $\pm$ one standard deviation from 10
    independent runs), and
    the lower panel [(c), (e)] shows the root mean squared scatter in transition
temperature values. 
    }

    \label{fig:walk_length_conv_peaks}
\end{figure}

As a basic test of the effectiveness of the three atomic motion
algorithms, SP-MC, GMC, and TE-HMC, we apply the NS method to a periodic Lennard-Jones system with interactions
truncated and shifted to zero at a cutoff distance of $3\sigma$
(see Appendix~\ref{app:LJpot}).
We choose a pressure above the triple point but below the
critical point, where this system
has two phase transitions: condensation from the gas to the
liquid, and freezing from the liquid to the crystalline solid.
Figure~\ref{fig:walk_length_conv_peaks} compares the performance of
the SP-MC, GMC and TE-HMC algorithms for resolving the heat capacity peaks
associated with these transitions.
The cost of each GMC timestep is similar to that in TE-HMC, and from Fig.~\ref{fig:walk_length_conv_peaks} we see that the convergence of all three methods is similar for the condensation transition.
However, TE-HMC is significantly more efficient than GMC for accurately resolving the freezing transition.

From Fig.~\ref{fig:walk_length_conv_peaks} we can see that, for equivalent accuracy, the number of all-particle sweeps required by the SP-MC algorithm is similar to the number of timesteps required by the TE-HMC and GMC algorithms.
For separable potentials such as LJ, an all-particle sweep takes roughly the same time as the full energy/force evaluation used in a timestep of GMC or TE-HMC. Therefore, for a non-separable potential, GMC and TE-HMC are $N$-times faster than the SP-MC algorithm.

The transition temperatures for the shorter, underconverged walks are
systematically lower than the fully converged value.  This underestimation is unsurprising:
if too few steps are taken at each iteration, then the sampler requires more 
iterations to find the low temperature phase.
Since subsequent NS iterations correspond to lower entropies and lower temperatures,
finding the structure at a later iteration corresponds to an underestimated
transition temperature.
The root mean square (rms) scatter in peak positions shown in Fig.~\ref{fig:walk_length_conv_peaks} does not go to zero for any 
of the methods, even with 
the longest walks  used ($L=2560$ energy/force evaluations):
for infinite walk lengths, the accuracy is limited by the number of walkers, $K$, and the number removed at each iteration, $K_r$.

\subsection{Example applications}\label{sec:ex_app}

In this section we demonstrate the utility of NS
by applying it to study four diverse systems: the order-disorder
transition of a binary LJ alloy, the eutectic of a copper-gold alloy,
the density anomaly of water which forms open crystal structures, and the phase
behavior of a bead-spring polymer model.  In all four cases we use TE-HMC to explore
the position degrees of freedom since it is most efficient for
this range of system sizes, as discussed below in Sec.~\ref{sec:results:PE_distribution}.
All simulations with the exception of the binary LJ were carried out with the LAMMPS package~\cite{LAMMPS}.

\subsubsection{Order disorder transition}

In addition to the condensation, freezing, and martensitic transitions that
have previously been simulated using NS~\cite{pt_phase_dias_ns}, multicomponent solids
also show transitions related to chemical ordering. 
Here we use NS to simulate the order-disorder transition of a model binary LJ alloy.
The potential energy function used, which favors the mixing of atoms, is given in Appendix~\ref{app:BLJpot}.

\begin{figure}
    \centerline{\includegraphics[width=0.9\columnwidth]{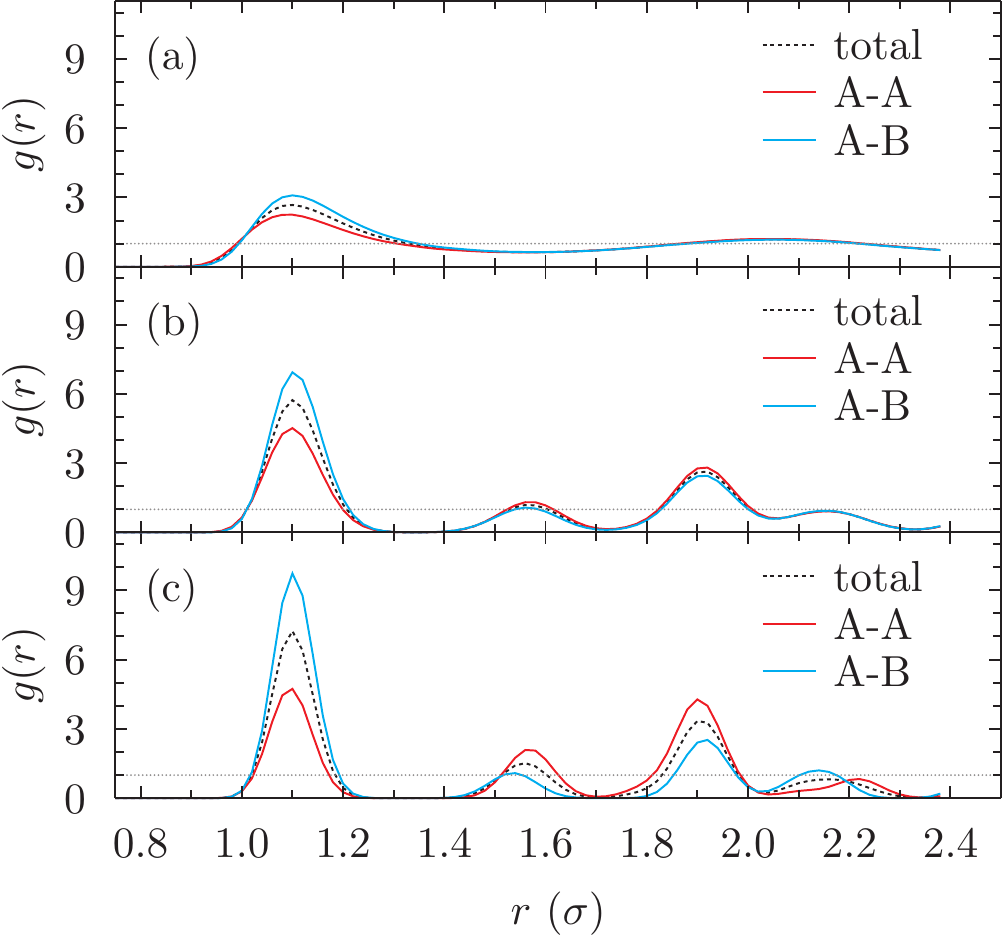}}
    \caption{ Ensemble-averaged radial distribution functions from NS
    phase diagram of binary LJ at equal composition. Panel (a)
    is above the melting point, (b) is between the freezing and chemical
    ordering transitions, and (c) is below the chemical ordering transition.
    Liquid (a) shows a low first neighbor peak and minimal other structure.
    Disordered solid (b) shows distinct peaks and some chemical
    ordering at first neighbors only.  Ordered solid (c) shows stronger
    chemical ordering, at least out to third neighbors.  
   These RDFs were calculated using a weighted sum of the RDFs
   of all configurations output by NS. 
   The RDF for each configuration was weighted by its Boltzmann weight $\left(\Gamma_{i-1} -\Gamma_i \right)e^{-\beta H_i}$, as in the partition function Eq.~\eqref{eq:hmc_ns_approx_npt_pf}.}
    \label{fig:binary_LJ_RDFs}
\end{figure}

\begin{figure}
    \centerline{\includegraphics[width=0.3\columnwidth]{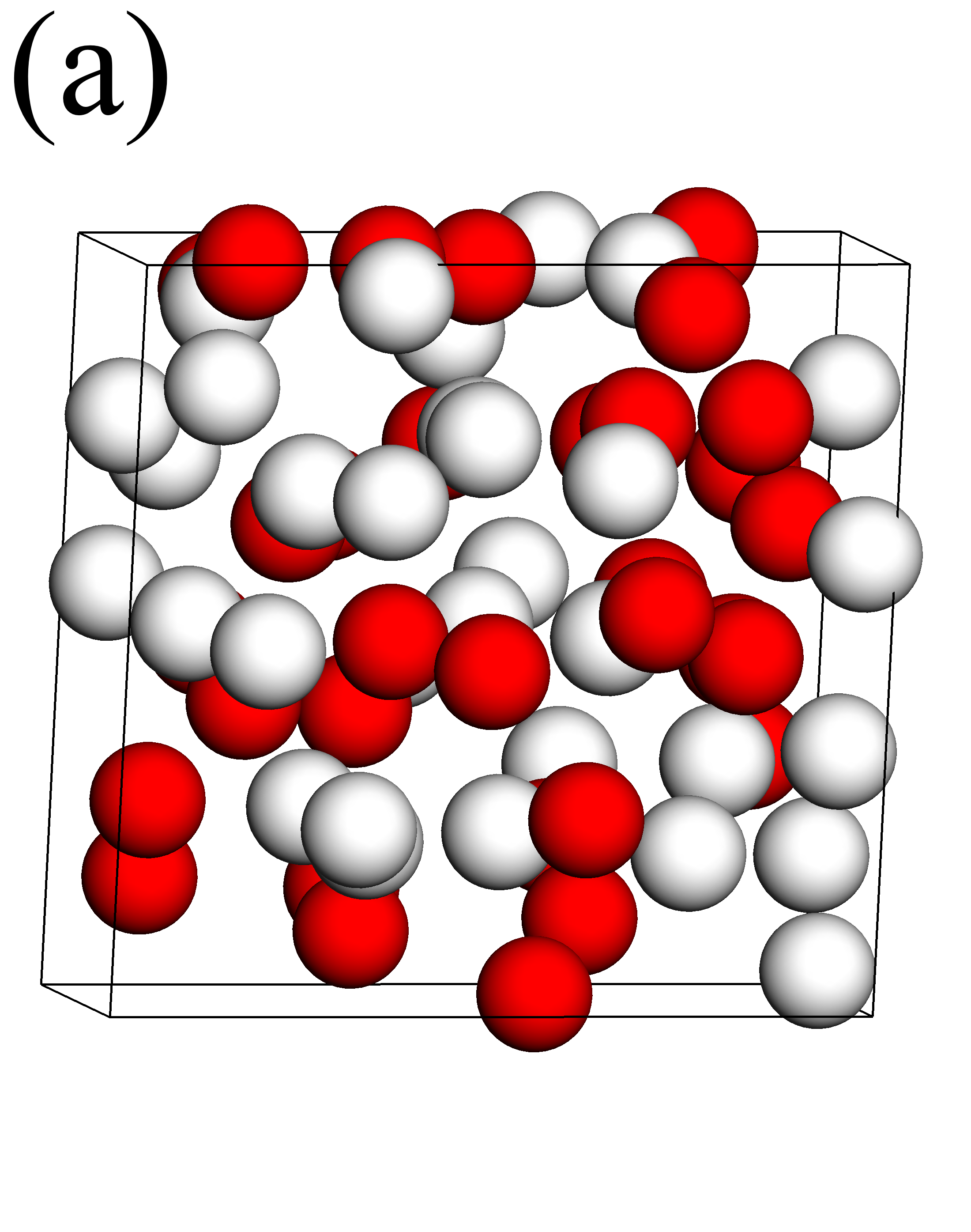}
                \includegraphics[width=0.3\columnwidth]{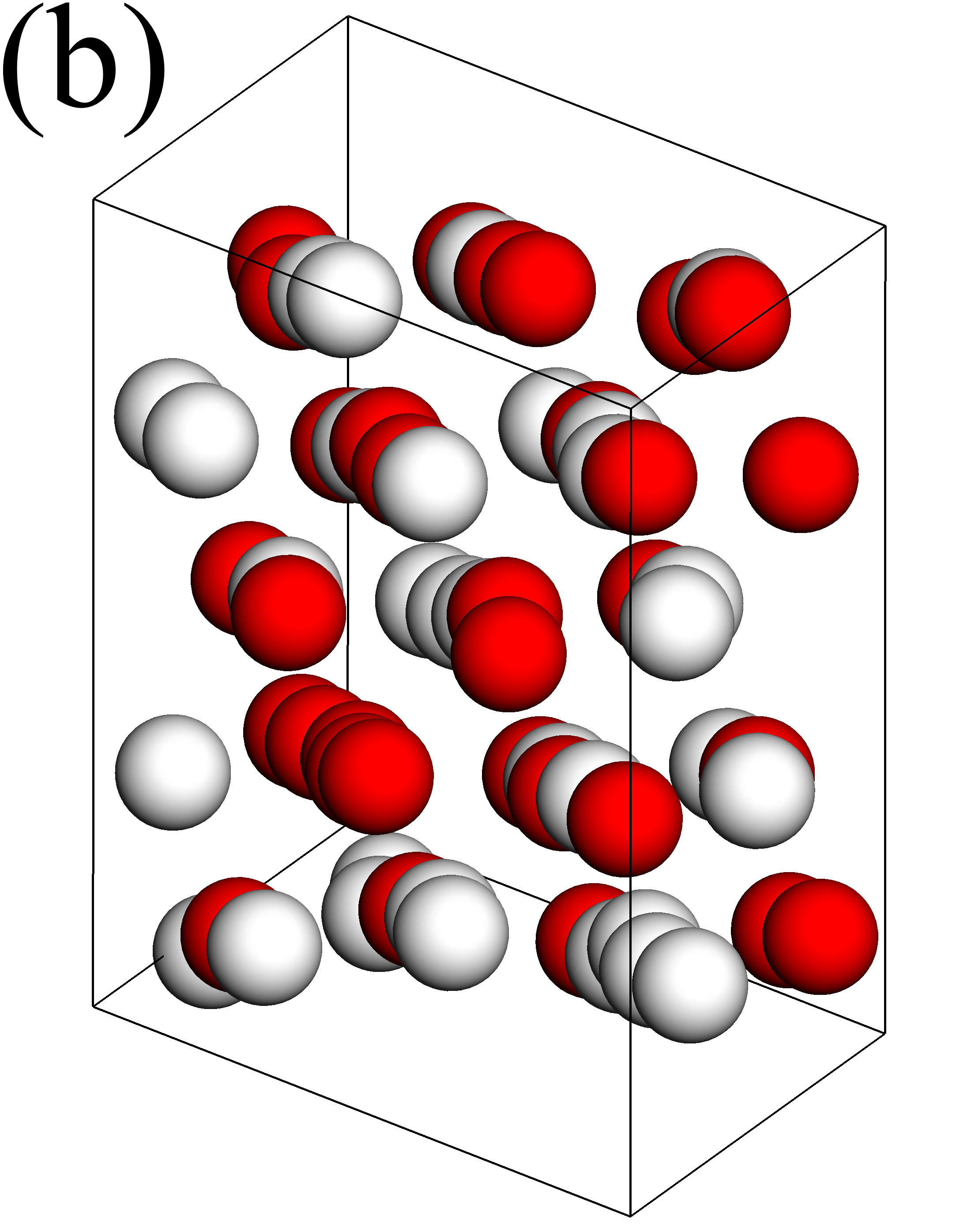}
                \includegraphics[width=0.3\columnwidth]{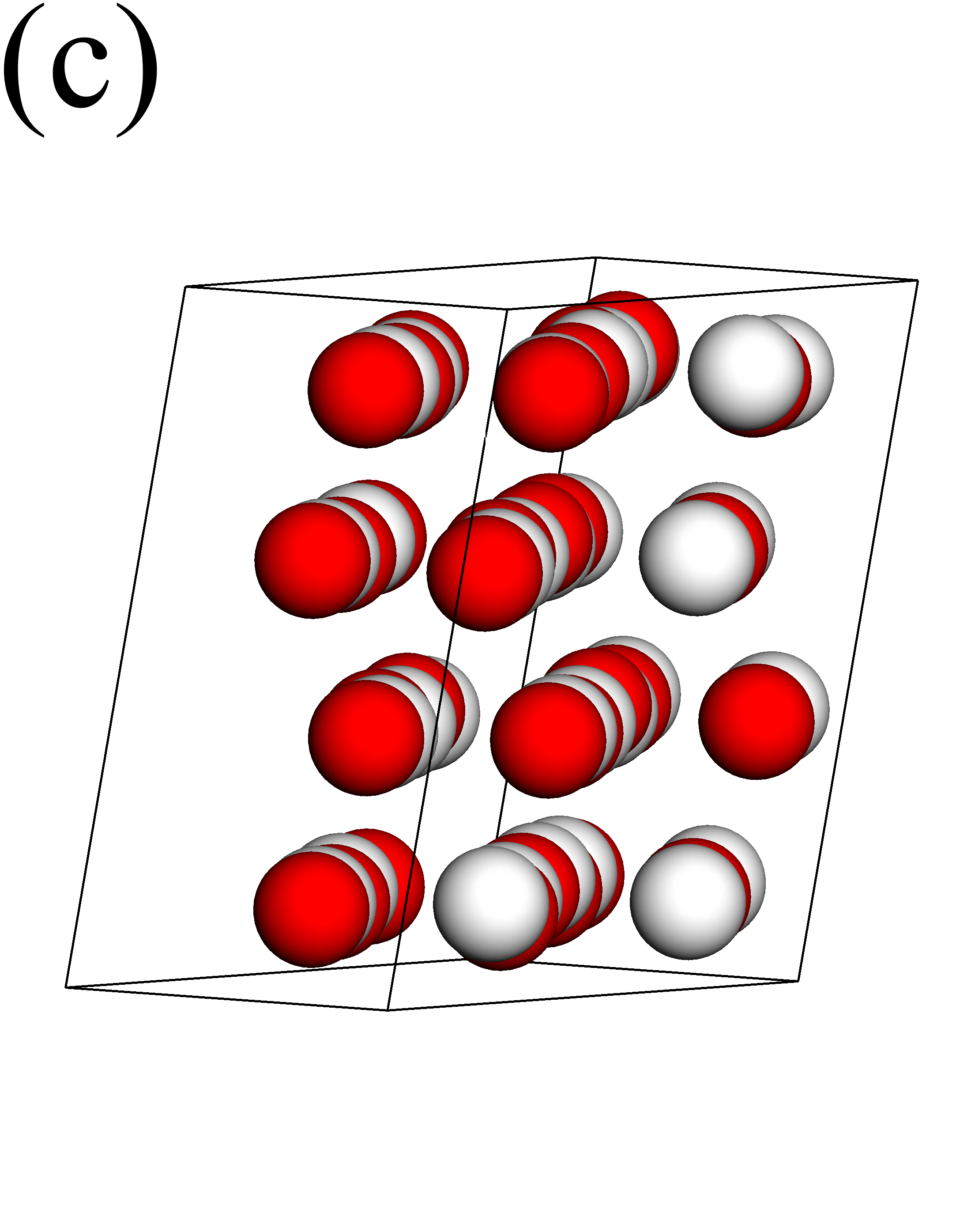}}
    \caption{Visualizations of typical configurations for
    the liquid phase [$T=0.95$, (a)], chemically disordered solid [$T=0.34$, (b)],
    and chemically ordered solid [$T=0.17$, (c)].}
    \label{fig:binary_LJ_pics}
\end{figure}

Figure~\ref{fig:binary_LJ_Cp} shows the heat capacity curve of this system.
The condensation, freezing and order-disorder transitions can be seen as three separate peaks.
Chemical ordering of the alloy can be observed in Figs.~\ref{fig:binary_LJ_RDFs} and~\ref{fig:binary_LJ_pics} which respectively show ensemble averaged radial distribution functions (RDFs) and typical configurations of the alloy at temperatures corresponding to the ordered and disordered solids, as well as the liquid.
Both the RDFs (Fig.~\ref{fig:binary_LJ_RDFs}) and the representative
atomic configurations (Fig.~\ref{fig:binary_LJ_pics}) show that
that at $T=0.95$, the system is a liquid, while at $T=0.34$ the system
is a chemically-disordered close-packed crystal, and at $T=0.17$ the crystal
is chemically ordered.

\begin{figure}
    \centerline{\includegraphics[width=0.9\columnwidth]{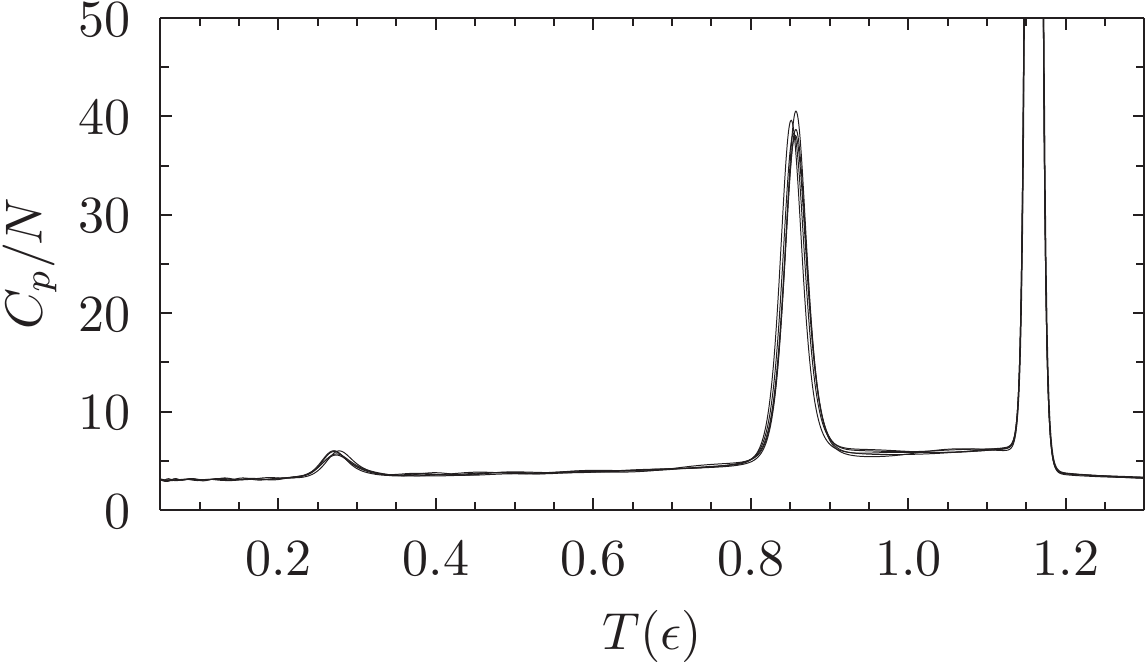}}
    \caption{Heat capacity as a function of temperature for a LJ binary alloy
    at equal composition for five independent NS runs.  The three peaks correspond to condensation,
    freezing, and chemical ordering, from high $T$ to low $T$.}
    \label{fig:binary_LJ_Cp}
\end{figure}

\subsubsection{Copper-gold eutectic}

\begin{figure}
    \centerline{\includegraphics[width=0.9\columnwidth]{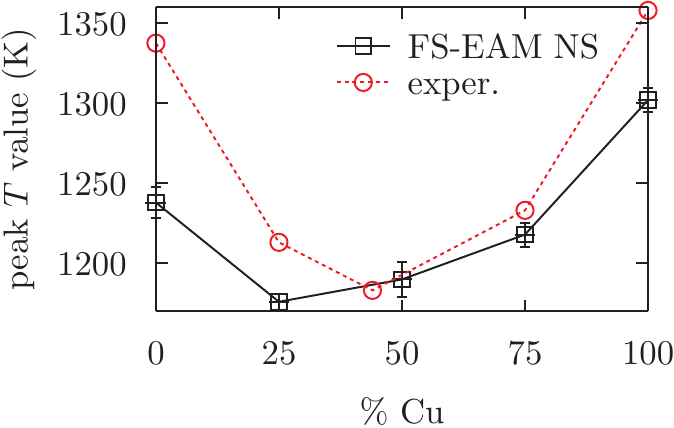}}
    \caption{Melting point as a function of composition in binary FS-EAM Cu$_x$Au$_{1-x}$~\cite{CuAu_FS_arxiv,CuAu_FS_web,Cu_Au_orig_FS}, calculated using the TE-HMC NS algorithm.
    Experimental values taken from~\cite{Okamoto}.
    Eutectic suppression of the melting point is observed at intermediate compositions.}
    \label{fig:eutectic_binary_FS}
\end{figure}

A eutectic, where the melting point of a multicomponent alloy is reduced at intermediate
compositions due to entropic effects, is an important example of the
interplay between energy and entropy affecting a phase transition.
We used NS to compute the heat capacity of
Cu$_x$Au$_{1-x}$ at a pressure of 0.1~GPa with interactions described by
a simple Finnis-Sinclair type embedded atom model (FS-EAM) inter-particle
potential~\cite{CuAu_FS_arxiv,CuAu_FS_web,Cu_Au_orig_FS} (see Appendix~\ref{app:cuaueam}).
In Fig.~\ref{fig:binary_FS_Cp} we show the heat capacity $C_p(T)$ for a number
of composition values $x$ at a temperature range near the melting point,
and in Fig.~\ref{fig:eutectic_binary_FS} we compare the resulting
computed melting points to experimental results from Ref.~\cite{Okamoto}.

\begin{figure}
    \centerline{\includegraphics[width=0.9\columnwidth]{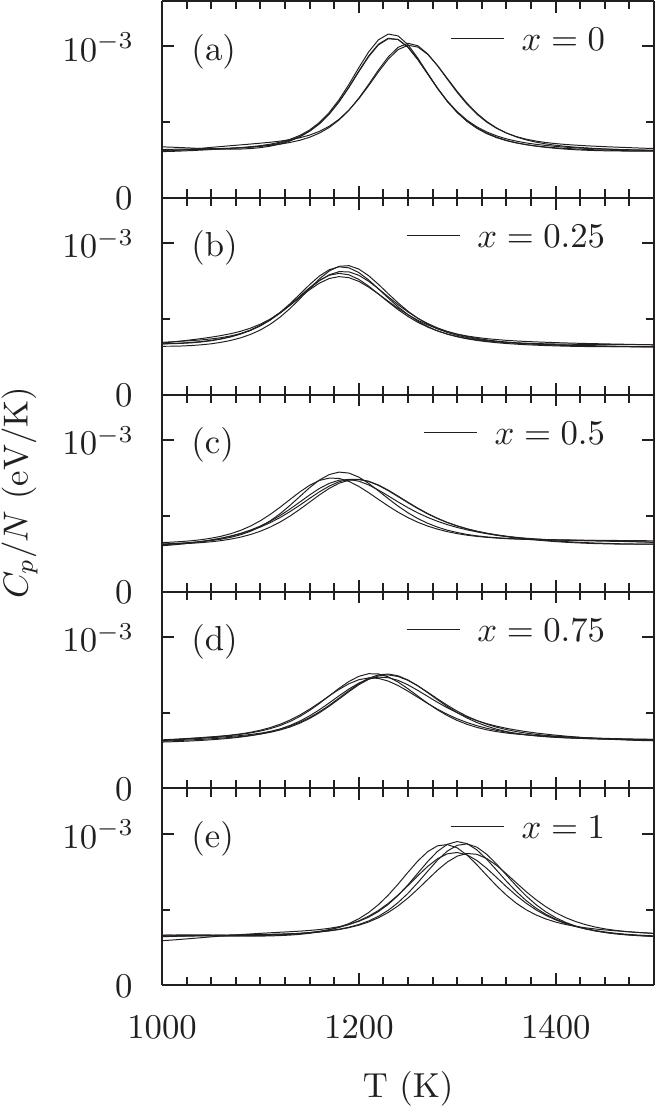}}
    \caption{Heat capacity as a function of temperature $C_p(T)$ of Cu$_x$Au$_{1-x}$ for a 
    range of compositions, from pure Au $x=0$ [panel (a)] to pure Cu $x=1$ [panel (e)]
    in 25\% intervals, showing variation of melting point peak. Each curve corresponds to an independent NS calculation.}
    \label{fig:binary_FS_Cp}
\end{figure}

We find that the FS-EAM potential gives a melting temperature $T_m \approx 1300$~K for pure Cu and $T_m \approx 1240$~K for pure Au.
The computed melting temperature is lower at all intermediate compositions, with a minimum value in the range 1175-1190~K between 25\% and 50\% Cu;
the computed melting temperatures at these two compositions are equal to within the error bars of the calculation.
While this FS-EAM potential underestimates the experimental melting points of the two endpoints, more severely so for Au (by 8\%), nested sampling shows that it reproduces the qualitative features of a eutectic.

\subsubsection{Density anomaly of water}

\begin{figure}
   \includegraphics[width=6.5cm,angle=90]{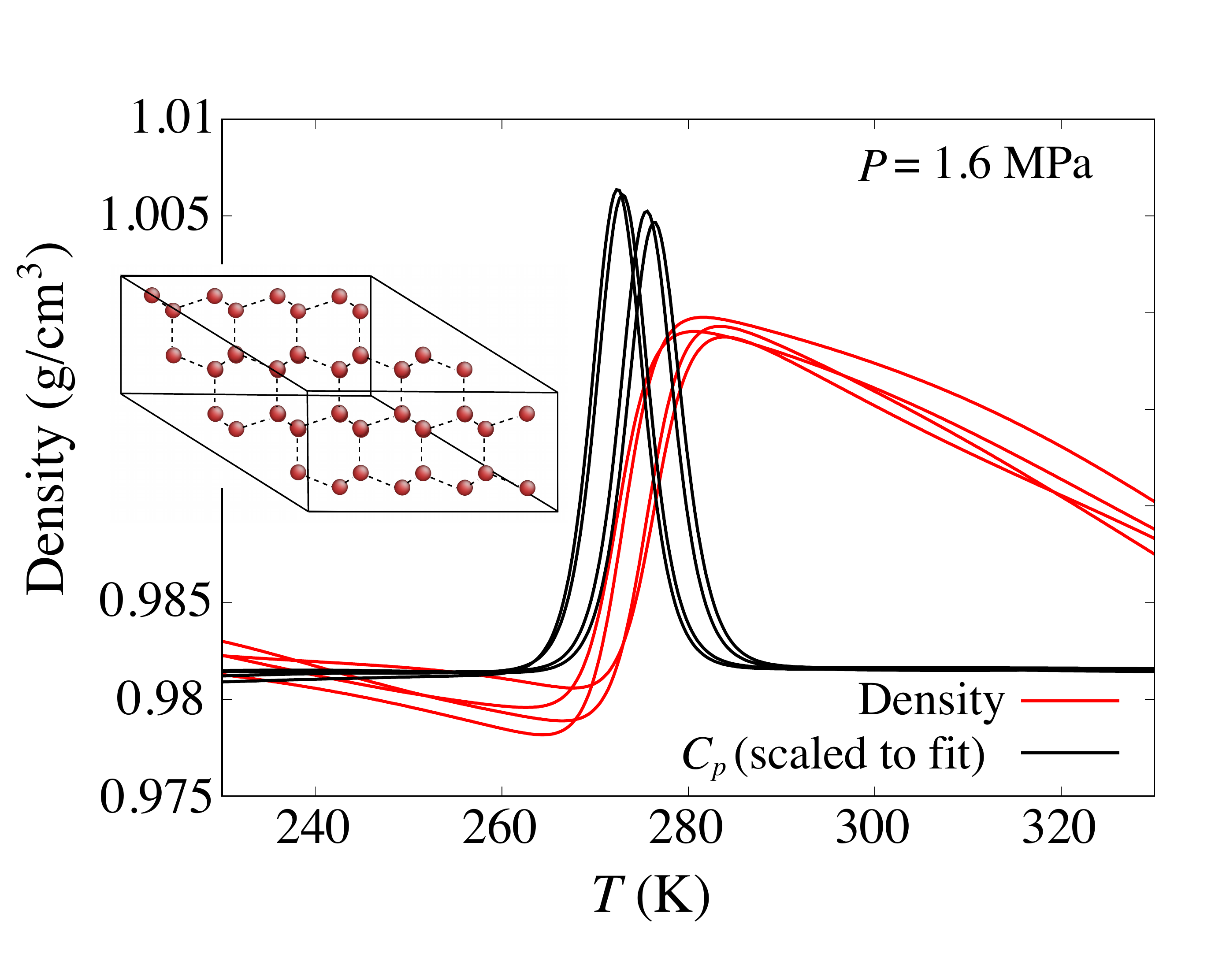}
   \caption{Heat capacity $C_p$ (in arbitrary units) and density curves for 64 mW water particles at a pressure of $1.6$~MPa, calculated using the TE-HMC NS algorithm. The inset shows a visualization of the hexagonal ice structure found by NS. Dashed lines in the inset represent the hydrogen-bond network.}   
   \label{fig:mW_water}
\end{figure}

The mW potential~\cite{mW_molinero} is a coarse-grained model of water,
designed to mimic its hydrogen bonded structure through a non-bonding
angular term, which biases the model towards tetrahedral coordination.
Despite only having short-range interactions and a single particle
representing a water molecule, it reproduces the energetics, anomalies,
liquid and hexagonal ice structure of water remarkably well. In order
to demonstrate that NS is capable of finding
not just close-packed but also open structures, we simulated 
water using  the mW potential. 
The computed heat capacity and density curves from four calculations  are shown in
Fig.~\ref{fig:mW_water}.  As expected, the particles were observed to
form a hexagonal ice structure, also shown in Fig.~\ref{fig:mW_water}.
By averaging the results from these calculations, we calculated the freezing temperature to be $274.3\pm 1.0$~K, the
density of ice to be $0.9792 \pm 0.0005~\mathrm{g/cm}^3$, and the density of the liquid at 298~K to be $0.9966 \pm 0.0004~\mathrm{g/cm}^3$. All these results
are in excellent agreement with values previously calculated for the mW
water model: $274.6$~K, $0.978~\mathrm{g/cm}^3$, and $0.997~\mathrm{g/cm}^3$, respectively~\cite{mW_molinero,mW_chandler}.  We found the maximum density of 
water to be $0.9992 \pm 0.0002~\mathrm{g/cm}^3$ at a temperature
$8.1\pm 0.3$~K above the freezing temperature. 

\subsubsection{Molecular solids \label{sec:res_mol_solid}}

Molecular materials are another system where nested sampling can be used to
efficiently sample the configuration space.  Both single-molecule systems and
multi-molecule systems, such as aggregating proteins or polymer melts, are of
interest.  In previous studies Wang-Landau sampling was used to map the
phase behavior of single polymer chains with different
lengths~\cite{Seaton.2010.PRE,Taylor.2009.JCP} and bending
stiffnesses~\cite{Seaton.2013.PRL, seaton2010wang}.
Here we present results for a bead-spring polymer model (see Appendix~\ref{app:beadspringpot})
with a harmonic bond and cosine angle potential parameterized by stiffnesses
$k_b$ and $k_a$, respectively, and a non-bonded LJ interaction with energy
$\epsilon$.  This model has been used to study crystallization in polymers~\cite{Karayiannis.2015}, and is similar to the model used in a previous
Wang-Landau study~\cite{Seaton.2010.PRE}.  

\begin{figure} 
\includegraphics[width=3.10in,angle=0]{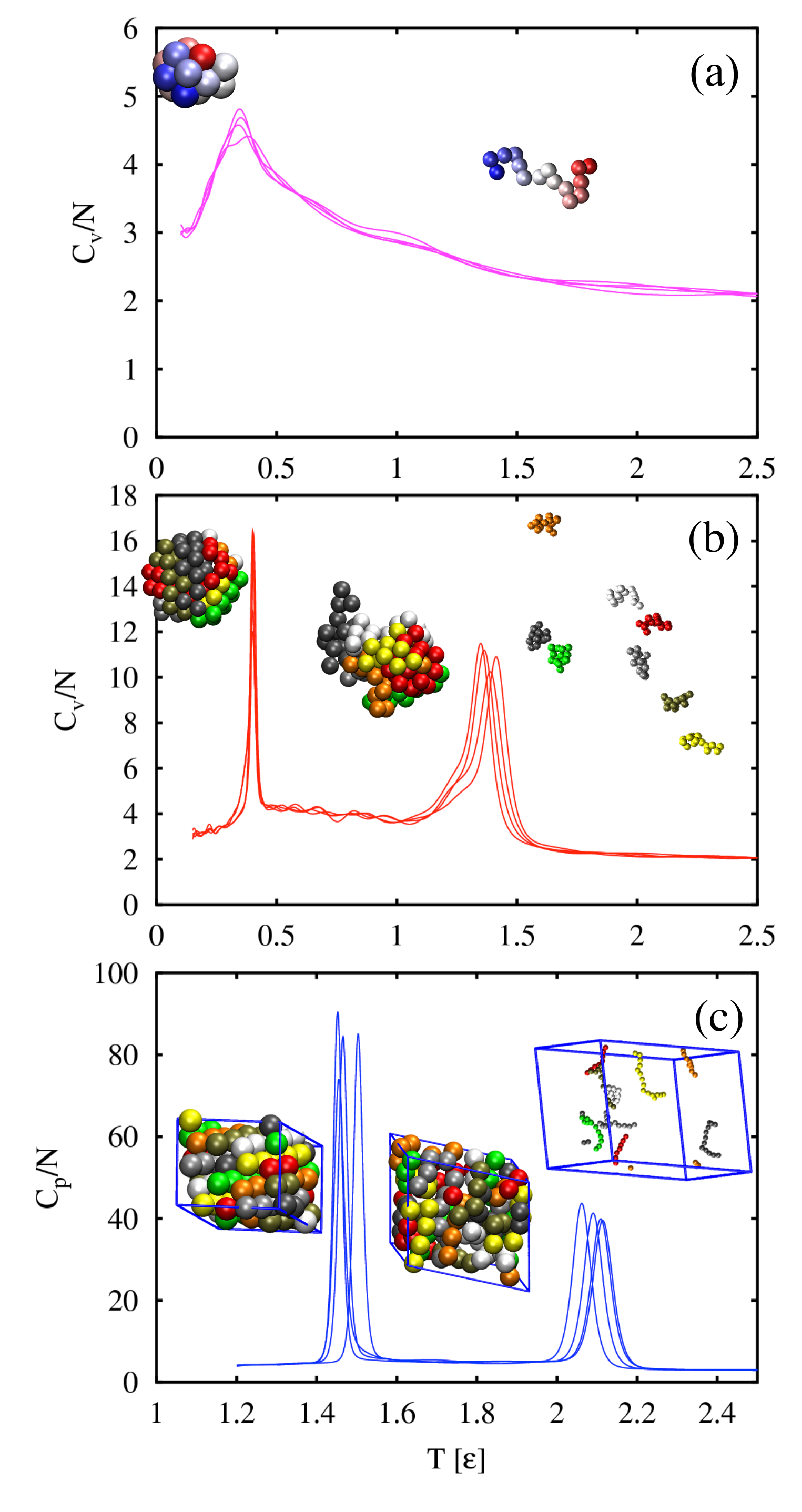} 
\caption{Heat capacity per coarse-grained bead  $C_p/N$ or $C_v/N$ curves for a single bead-spring
polymer chain of 15 beads with monomer density $2.5\times10^{-5}\sigma^{-3}
\approx \left(\frac{1}{40\sigma}\right)^3$~[panel (a)], 8 15-bead chains in a
periodic box with monomer density $2\times10^{-3} \sigma^{-3} \approx
\left(\frac{1}{8\sigma}\right)^3$~[panel (b)], and 8 15-bead chains in a periodic
box with cell moves and $P=2.3\times10^{-3}\epsilon/\sigma^{3}$~[panel (c)].
Both constant-volume systems use fully flexible ($k_a = 0$) models, while the
constant pressure system (bottom) has $k_a = 10\epsilon.$  Snapshots show
example polymer conformations corresponding to the different phases.}
\label{f:polymer-hc} \end{figure}

Figure~\ref{f:polymer-hc} shows heat capacity curves calculated for three
different systems: (i) a single, fully flexible ($k_a = 0$) bead-spring
polymer chain of 15 beads in a constant-volume periodic box with monomer
density $2.5\times10^{-5}\sigma^{-3} \approx \left(\frac{1}{40\sigma}\right)^3$;
(ii) 8 fully flexible 15-bead chains in a constant-volume periodic box with
monomer density $2\times10^{-3} \sigma^{-3} \approx
\left(\frac{1}{8\sigma}\right)^3$; (iii) 8 15-bead chains with angle stiffness
$k_a = 10\epsilon$ at fixed pressure $P=2.3\times10^{-3}\epsilon/\sigma^{3}$,
and flexible periodic boundary conditions.  The low monomer densities of the
constant-volume systems do not allow for a single chain to interact with itself
through the periodic boundary at the temperatures of interest.  Snapshots of 
configurations for each system type are also shown, illustrating the observed
phase transitions.  

The single chain shows a broad transition below $T=0.5~\epsilon$, from an
extended state to a collapsed, ordered state, in agreement with previous
results~\cite{Seaton.2010.PRE}.  
For short, unentangled polymer chains nested sampling leads to   slightly
lower relative error in the peak height when compared to previous
results at approximately half the computational cost~\cite{seaton2010wang, Seaton.2010.PRE}.
The constant-volume multichain system has two
transitions: first, chain aggregation occurs at $T=1.4~\epsilon,$  and second,
at $T=0.4~\epsilon$ the monomers order, forming a solid cluster with high-symmetry. 
This is reminiscent of the  $N=100$ single-chain transition observed
previously~\cite{Seaton.2010.PRE}.  The multichain periodic system has two
transitions, the first at $T=2.05~\epsilon$ from a polymer gas to a melt, and the
second at $T=1.45~\epsilon$ from a melt to a crystalline solid, in agreement
with MD simulations of polymer crystal nucleation~\cite{Karayiannis.2015}.  

\subsection{System size dependence of enthalpy distribution}
\label{sec:results:PE_distribution}

As mentioned in Sec.~\ref{sec:MCMCsteps:GMC}, nested sampling using GMC creates
a series of probability distributions~\eqref{eq:fixedP_ns_dist}
that correspond to  uniform distributions in the  Cartesian
particle coordinates, $\mathbf{r}$, such that $\widetilde{H}(\mathbf{r}) <
\widetilde{H}^\mathrm{sup}$.  In this case $\mathrm{Prob}(\widetilde{H})
$ is proportional to the density of states for $\widetilde{H}$,
which is strictly unimodal.  In contrast, TE-HMC works by performing
nested sampling in total phase space, and samples from a series
of probability distributions~\eqref{eq:fixedP_ns_dist_tehmc} that
correspond to uniform distributions in the  phase space
coordinates $(\mathbf{r},\mathbf{p})$, with $H(\mathbf{r},\mathbf{p})<
H^\mathrm{sup}$.  In TE-HMC, the marginal distribution for $\mathbf{r}$
is {\em\ not} a uniform distribution, and for larger system sizes a bimodality
is observed in the probability distribution for $\widetilde{H}$ at
phase transitions.

\begin{figure}
    \centerline{\includegraphics[width=\columnwidth]{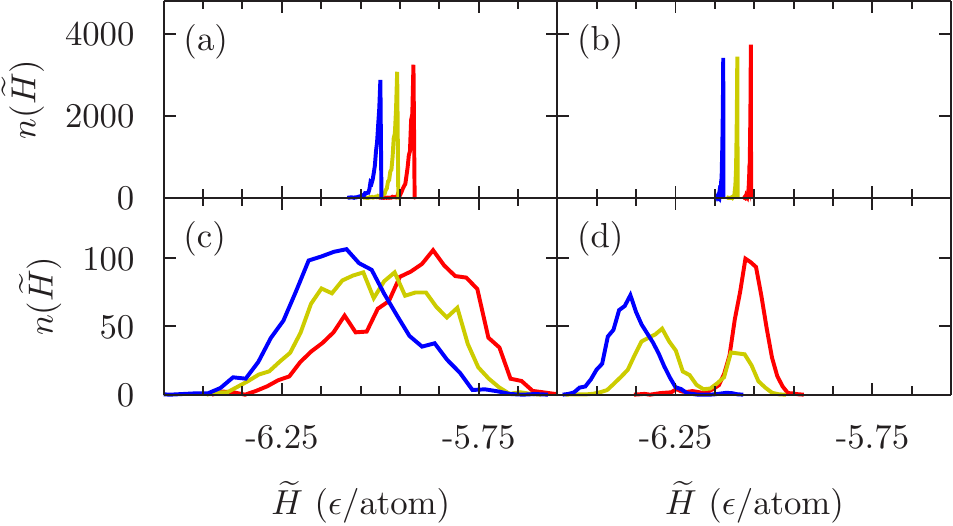}}
    \caption{Distribution of configurational enthalpy  (excluding kinetic
    energy contribution for TE-HMC)
    for a 64 atom system [left column (a), (c)]
    and 256 atom system [right column (b), (d)] of monatomic LJ, for a range
    of NS iterations that spans the highest weight
    configurations for the freezing transition at $T \sim 0.65~\epsilon$.
    Top row [(a), (b)] shows GMC results and bottom row [(c), (d)] shows TE-HMC.  Colors
    indicating iteration, from earliest (highest energy) to latest (lowest energy) are red,
    yellow, and blue.
    }
    \label{fig:PE_distribution} 
\end{figure}

Figure~\ref{fig:PE_distribution} compares the observed  probability
distributions for $\widetilde{H}$ in the region of the freezing
transition (for the same pressure as the system presented in Sec.~\ref{sec:results:mono_LJ_walk_length})
with the TE-HMC and GMC algorithms, for simulations
of 64 particles (as used in the earlier subsection) and also for 256 particles.
Using TE-HMC for 64 particles
$\mathrm{Prob}(\widetilde{H})$ is unimodal and broadens slightly at
the freezing transition, but never becomes bimodal.  For the larger 256
particle system, $\mathrm{Prob}(\widetilde{H})$ becomes bimodal, which can be
clearly seen in the middle curve (yellow).

In order to obtain an accurate estimate of the integrated density of states, $\Gamma(H)$, TE-HMC must draw a  sample from the uniform distribution in $(\mathbf{r},\mathbf{p})$, Eq.~\eqref{eq:fixedP_ns_dist_tehmc}, at each iteration.
NS approaches each transition from above, and if $\mathrm{Prob}(\widetilde{H})$ is bimodal, initially all $K$ configurations will  be in the mode at higher $\widetilde{H}$.
To draw a proper sample from $\mathrm{Prob}(\widetilde{H})$ at the phase transition, the MCMC walk must be long enough that the configuration can feasibly pass back and forth between the two modes, traversing the intermediate range of $\widetilde{H}$, several times.
For larger $N$, as  $\mathrm{Prob}(\widetilde{H})$  gets smaller in the region between the modes,  transitions between the two modes will become less frequent and much longer MCMC trajectories will be required at iterations close to the phase transition.

We observed in Sec.~\ref{sec:results:mono_LJ_walk_length} that, for
simulations of 64 particles, TE-HMC is significantly more efficient
than GMC for accurately resolving the freezing transition.  However,
as a result of the bimodality in $\mathrm{Prob}(\widetilde{H})$, GMC
may become more efficient than TE-HMC for larger system sizes.
In the future, it would be desirable to develop algorithms which, like
TE-HMC, use atomic forces at every step to expedite configuration space
exploration, yet avoid this bimodality in $\mathrm{Prob}(\widetilde{H})$
at larger system sizes.

\section{Conclusions}

In this paper we have proposed efficient all-particle moves using inter-particle forces and dynamics 
 for constant pressure nested sampling.  The TE-HMC, GMC and SP-MC algorithms reach the same accuracy 
 using approximately the same
number of full system energy evaluations (TE-HMC and GMC) or full SP-MC sweeps.  
For separable potentials, where a single particle move can
be computed in $1/N$ the cost of a full system energy evaluation,
this makes the three methods equally efficient; for non-separable potentials,
where such efficient single atom moves are not possible, the TE-HMC
and GMC algorithms are $N$ times faster.

The TE-HMC algorithm uses constant energy molecular dynamics,
implemented in many software packages, but requires extending the
NS method to sample positions and momenta, and leads to increasingly bimodal
configurational enthalpy distributions as the system size increases.
This bimodality is likely to make equilibration and sampling difficult
for sufficiently large systems, although this has not been a practical
problem for the 64-120 particle systems we have considered here.
The GMC algorithm, although somewhat less efficient in this size
range, maintains the unimodal configurational enthalpy distribution
of the previous Gibbs-sampling-based approach, and is therefore not
expected to suffer from a breakdown in ergodicity for larger systems.

We have implemented the constant pressure NS method using these
algorithms in the {\tt pymatnest} software~\cite{pymatnest}, which includes a parallel
algorithm and a link to the {\tt LAMMPS} package which itself has many
inter-particle potentials available. Using this implementation we have
shown that the constant pressure NS method with these algorithms can
be used to simulate a wide range of systems with different interaction
potentials and types of phase transitions: order-disorder transitions
in binary LJ, eutectic composition dependence of the melting point
in Cu-Au, freezing of water which has a density anomaly and an open
crystal structure, and condensation and solidification of a bead-spring polymer
model.

\begin{acknowledgments}

R.J.N.B. acknowledges support from EPSRC Grant No. EP/J017639/1.
G.C. acknowledges support from EPSRC under Grants No. EP/P022596/1 and No. EP/J010847/1.
L.B.P. acknowledges support from the Royal Society.
The work of N. B.
was supported by the Office of Naval Research through the U.~S. Naval Research
Laboratory's 6.1 base program.  K. M. S. was supported
by the National Research Council's Research Associateship Program 
at the U.~S. Naval Research Laboratory.
N.B. and K.M.S.
acknowledge computational resources from the
DOD
High Performance Computing Modernization Program Office
(HPCMPO) at the ARL and AFRL DSRCs.

\end{acknowledgments}

\appendix

\section{NS stopping criteria: temperature estimates
\label{app:Testimate}}

The simplest criteria for stopping the NS iteration are 
to use a fixed number of iterations (i.e.\ fixed reduction in entropy)
or a fixed potential energy minimum (e.g. close to the ground state).  A more
physically appealing criterion takes advantage of the approximate
correspondence between the downward scan in enthalpy and
decreasing temperature.
The estimate we use to terminate the outermost NS iteration
loop is based on the expressions for thermodynamic quantities, such
as the partition function (or enthalpy or heat capacity, which are its
derivatives).  We find that the range of iterations that contribute
with significant weight to the ensemble average at each temperature is
sharply peaked.  When the contribution of the current iteration to the
partition function at a specified temperature $T_\mathrm{min}$ is a factor
of $e^{10}$ lower than the maximum contribution of any previous iteration,
we assume that no later iteration will contribute significantly, and therefore
consider the calculation to be converged for all $T \ge T_\mathrm{min}$.
We use this stopping criterion in all simulations reported here. Note
that a monotonic relationship between iteration and temperature is
not always satisfied; near phase transitions the dependence is more
complicated, and setting $T_\mathrm{min}$ too near a phase transition
will lead to unpredictable behavior.  

The convergence criterion described above is efficient enough to evaluate
at each iteration for a single choice of $T_\mathrm{min}$, but too
computationally expensive to use as an estimate of the ``current temperature''
during NS iterations, because it would need to be evaluated for many
values of $T$ to find the lowest.  We therefore use an independent
estimate of the temperature to monitor the progress of the NS iterations.
This estimate is based on the rate of decrease of $\widetilde{H}^\mathrm{sup}$
(or $H^\mathrm{sup}$)
as a function of iteration number.
The iteration number $i$ is linearly related
to the logarithm of configuration space volume (i.e.\ microcanonical entropy $S$), and
that rate of decrease is therefore related to $\partial \widetilde{H}/\partial S$.
The current temperature during the NS simulation can be estimated from the finite difference expression
\begin{equation}
    T \approx \left( k_B \frac{D \ln \alpha}{\widetilde{H}^\mathrm{sup}(i-D)-\widetilde{H}^\mathrm{sup}(i)}  \right)^{-1} \label{eq:T_est}
\end{equation}
where $D$ is an interval over which the finite difference is taken (1000
iterations here), and $k_B$ is Boltzmann's constant.  
We use this expression to monitor the progress of the NS iterations,
but not to terminate.

\section{Parallelization}
\label{app:para}

The \texttt{pymatnest} software~\cite{pymatnest}, in which these algorithms
are implemented, combines two separate forms of parallelization
which were previously reported separately. In~\cite{pt_phase_dias_ns},
during step 2 of the NS algorithm, rather than decorrellating a single cloned
configuration alone using a MCMC walk comprising $L$
energy evaluations, the authors evolve $n_p$ 
configurations (including the cloned configuration) in parallel using $n_p$ processes through $L/n_p$
MC moves each.  Each configuration is evolved for an average
of $n_p$ iterations before being recorded and removed.
Thus the user specifies
the {\em average} number of energy evaluations used to decorrellate a cloned
configuration from its starting coordinates.  
In~\cite{Burkoff1, bib:Frenkel_NS}, on the other hand, the authors evolve each cloned configuration for exactly $L$ steps,
but they parallelize over $n_p$ processes by removing $K_r = n_p > 1$ configurations at
each NS iteration, resulting in $K_r$ cloned configurations that can be walked
in parallel.

In \texttt{pymatnest}, we combine the two formulations by allowing for
$K_r > 1$ and evolving the $K_r$ cloned configuration in parallel, but
also for the number of parallel tasks $n_p > K_r$, reducing the walk length required
at each iteration.
To optimize load
balance, each of the $K_r$ cloned configurations that must be evolved
is assigned to a different parallel task, and all remaining parallel
tasks (which would otherwise be idle if $n_p > K_r$) walk $K_e = n_p-K_r$ additional
randomly chosen configurations.

From the probabilities for a configuration to be removed or walked at
each NS iteration it is possible to calculate the distribution 
of the number of walks each configuration has experienced,
and from that the mean number of times a configuration will be walked before it is 
removed $\langle n_\mathrm{walks} \rangle$.  
The 
general expression for the length of 
the walk that must be done at each iteration to achieve an expected total walk 
length $\langle L \rangle$, for arbitrary $K$, $K_r$, and $K_e$ is 
\begin{equation} \label{eqn:walk_len} 
L' = \langle L \rangle / \langle n_\mathrm{walks} \rangle  = \langle L \rangle \frac{K_r}{K_r+K_e} = \langle L \rangle \frac{K_r}{n_p}.
\end{equation}

For each MCMC walk in \texttt{pymatnest}, different move types are randomly 
chosen from the list of possible moves with predetermined ratios
until at least $L'$ energy evaluations have been performed.

To maintain load balance
the shortest walk must be a few times longer than the longest possible
single step, for example, a single SP-MC sweep or an MD/GMC
trajectory.  Therefore the maximum parallelization, $n_p$, that can be achieved
depends only on the total walk length, $L$, and number of configurations removed at each
iteration, $K_r$, and not on the number of configurations, $K$.  For typical
runs we show here, $K_r=1$, $L \approx 500-1000$, and the length of
each MD trajectory is 8.  To keep reasonable parallel efficiency we find
that $L'$ must be larger than about 20, so the maximum $n_p \approx 25-50$.  

\subsection{Qualitative behavior of parallelized NS}

The computational cost and accuracy of NS depend on these parameters
in a complex way.  The total computational work is proportional to $K$ and
$\langle L \rangle$, and is independent of $K_r$ and $n_p$.  If $K$ is
increased at fixed $K_r$, the fraction of configuration space that remains
after each iteration $\alpha = (K+1-K_r)/(K+1)$ comes closer to 1.0,
and the number of NS iterations required increases approximately linearly
with $K$.   If instead $K$ and $K_r$ are increased proportionately,
$\alpha$ and therefore the number of iterations remain roughly constant,
but the work at each iteration (to walk $K_r$ cloned configurations)
increases proportionately.  It is not clear {\it a priori} how the
necessary value of $\langle L \rangle$ changes with $K$: there is some evidence
that once $K$ is large enough, increasing it further reduces
the distance each cloned configuration must be walked to decorrelate
it sufficiently, but this relationship requires further investigation.

The accuracy of the configuration space volume estimates computed by NS
also depends on $K$ and $K_r$.  The value of $\alpha$ determines the
resolution in configuration space volume, but larger values of $K$
(at constant $\alpha$) reduce the {\em noise} in the estimate (for the
same reason that the 500th sample out of 999 is a less noisy estimator of
the median than the 2nd sample out of 3).  The value of $\langle L \rangle$
also affects the error, because insufficiently walked configurations
have a correlation to the configuration they were cloned from, which leads
to a deviation from the uniform distribution.

Since the useful parallelism is limited by the minimum value of $L'$,
which is clearly independent of $K$, only increasing $\langle L \rangle$
or $K_r$ can increase it.  The former is useful only up to the point
where the configurations are sufficiently decorrelated, as our convergence
plots in Sec.~\ref{sec:results:mono_LJ_walk_length} show.  Increasing the latter at constant $K$ decreases
the resolution in configuration space volume (by decreasing $\alpha$),
and therefore leads to increased error. 
Increasing $K_r$ while also
increasing $K$ proportionately maintains the resolution and actually
reduces the noise in the configuration space volumes, but also increases
the computational work, but not necessarily the time to solution if $n_p$
can also be increased proportionately.

\subsection{Quantitative behavior of parallelized NS}

In this section we limit the discussion to the original constant
pressure, flexible periodic boundary conditions nested sampling
algorithm as implemented by SP-MC and GMC.  The same discussion can
be extended exactly to TE-HMC by a change of symbols: $H$ for
$\widetilde{H}$ and $\Gamma$ for $ \chi$ (see Sec.~\ref{sec:MCMCsteps}),
so long as one takes care not to confuse the phase space volume
$\Gamma$ and the gamma function in Appendix~\ref{sec:para_kr_gt1}.
In the next two subsections we separately discuss the effect of the two 
approaches to parallelization: first varying $K_r$ while assuming our MCMC draws perfect samples from the probability distribution~\eqref{eq:fixedP_ns_dist} [or~\eqref{eq:fixedP_ns_dist_tehmc}];
 second varying $K_e$ at fixed $\langle L \rangle$ and $K_r$. 

\subsubsection{$K_r>1$ and $K_e=0$ \label{sec:para_kr_gt1}}

In this subsection, we assume that our MCMC walk yields perfect samples from the distribution~\eqref{eq:fixedP_ns_dist} (or~\eqref{eq:fixedP_ns_dist_tehmc}).
In step 1 of the NS algorithm (see Sec.~\ref{sec:method}) $\widetilde{H}^\mathrm{sup}$ is updated to the lowest of the $K_r$ highest enthalpies in our sample set, and the configuration space volume contained by the updated $\widetilde{H}^\mathrm{sup}$ is $\chi_i\approx \chi_0[(K-K_r+1)/(K+1)]^i$, where $i$ is the NS iteration number.
For $K_r>1$ it is also possible to give  analytic estimates of the configuration space volumes contained by the $K_r-1$ higher enthalpy values between $\widetilde{H}^\mathrm{sup}_{i-1}$ and $\widetilde{H}^\mathrm{sup}_i$~\cite{bib:Frenkel_NS}.
Thus one may consider the configurational entropy contained at fractional numbers of NS enthalpy levels.

After a number of enthalpy levels
\begin{equation} \label{eq:n_logD}
n_{\Delta} =  \left(\sum_{i=K-K_r+1}^{K}{\frac{1}{i}} \right)^{-1}  
\end{equation}
the expectation of the logarithm of the configuration space enclosed by $\widetilde{H}$  decreases by 1: 
\begin{equation} \label{eq:mean_logD}
\langle \ln\chi_i - \ln\chi_{i+n_\Delta} \rangle = -1.
\end{equation}
If we assume that it is possible to draw perfect random samples from~\eqref{eq:fixedP_ns_dist} (or~\eqref{eq:fixedP_ns_dist_tehmc}), then it can be shown that, after the same number of enthalpy levels $n_{\Delta}$, the variance of $\Delta\ln\chi=\ln\chi_i - \ln\chi_{i + n_\Delta }$ is given by
\begin{equation} \label{eq:var_logD}
\mathrm{Var}(\Delta\ln\chi) = \frac{d^{(K-K_r+1)}\Gamma(z)}{dz^{(K-K_r+1)}}\Bigg|_{z=1} - \frac{d^{(K+1)}\Gamma(z)}{dz^{(K+1)}}\Bigg|_{z=1}
\end{equation}
where $\Gamma(z)$ is the gamma function.
The standard deviation, $\left[\mathrm{Var}(\Delta\ln\chi)\right]^{\frac{1}{2}}$, represents the rate at which uncertainty in $\ln\chi$ accumulates during a nested sampling calculation.
For a serial calculation ($K_r=1$ and $K_e = 0$), $\left[\mathrm{Var}(\Delta\ln\chi)\right]^{\frac{1}{2}} = \frac{1}{\sqrt{K}}$.

\begin{figure}
    \centerline{\includegraphics[width=0.9\columnwidth]{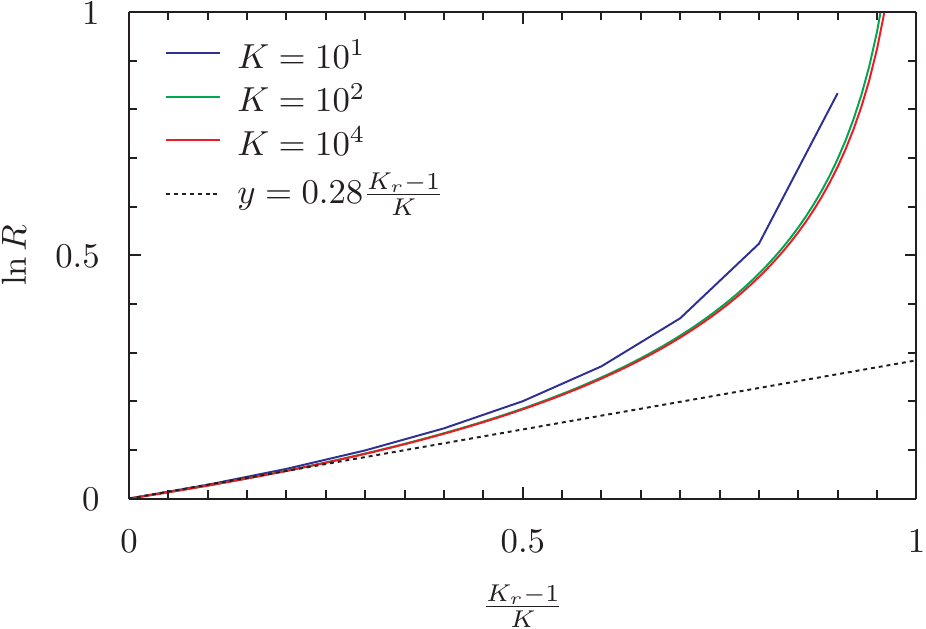}}
    \caption{Ratio of the rates, $R$,  at which uncertainty in $\ln\chi$ accumulates for runs
    parallelized by removing and walking $K_r > 1$ as compared to serial run with $K_r = 1$,
    as a function of scaled number of configurations removed $\frac{K_r-1}{K}$, for several values of $K$.  
    Dashed line indicates linear trend for $K_r -1 \ll K$.}
    \label{fig:rel_error_discard_kr}
\end{figure}

Figure~\ref{fig:rel_error_discard_kr} shows how the ratio of $\left[\mathrm{Var}(\Delta\ln\chi)\right]^{\frac{1}{2}}$ for parallel and serial NS, $R = [\mathrm{Var}(\Delta\ln\chi)]^{\frac{1}{2}}\div\frac{1}{\sqrt{K}}$, depends on $\frac{K_r}{K}$.
$R$ represents the relative rate at which uncertainty in $\ln\chi$ accumulates during parallel and serial calculations.
One can  see that  $\ln{R}$ converges for $K>10^2$, and for $\frac{K_r}{K}\lesssim 0.25$, $ \left[ \mathrm{Var}(\Delta\ln\chi)\right]^{\frac{1}{2}} \sim \exp{\left(0.28 \frac{K_r-1}{K}  \right)} \frac{1}{\sqrt{K}}$.
For larger $\frac{K_r -1}{K}$, $R$ increases  more rapidly.

\subsubsection{$K_e > 0$ and $K_r = 1$}

Parallelizing with $K_r = 1$ by using additional tasks to walk $K_e > 0$ extra
configurations (in addition to those that were cloned) changes the walk
length $L$ from a deterministic parameter of the NS method to a stochastic one.
It is possible
to derive the variance of the walk length
\begin{equation}
\mathrm{Var}(L) = \langle L \rangle^2\, \frac{K_e  }{ K_e+K_r}
\end{equation}

While unlike the case of $K_r > 1$ there are no analytic results for the error due to the variability of $L$, 
several observations can be made.  One is that the square root of the variance (except for the serial case of 
$K_e = 0$) is almost as large as $L$ itself.  Another is that the scaling of $\mathrm{Var}(L)$ with the 
number of extra parallel tasks $K_e$ is polynomial, unlike the exponential scaling of the error $R$ with 
extra tasks for $K_r > 1$.  It is unclear, however, how this variability in walk length will affect the
error in the results, although the  empirical observation is that this effect does not appear to be strong.  
Nevertheless, the serial case of $K_r = 1$ and $K_e = 0$ results in the lowest uncertainty in estimates of 
configuration or phase space volumes.

\begin{alg} [caption={[null] Total enthalpy Hamiltonian Monte Carlo step for nested sampling.   Repeated application converges to uniform sampling of phase space in the region $H<H^{\mathrm{sup}}$, $E_k(\mathbf{p})<E_k^0$ . },label=alg:ns_hmc]
subroutine te_hmc$(\mathbf{s},\mathbf{p},V,\mathbf{h}_0, M, dt)$
  ! Nested sampling TE-HMC step
  
  $ E_{k}^{\mathrm{max}}=\min\left[E_k^0, H^{\mathrm{sup}}-\widetilde{H}\left( \mathbf{s},V,\mathbf{h}_0 \right) \right] $
  
  if ( complete_momentum_rand ):
    $\mathbf{p}$ = complete_rand_p($E_{k}^{\mathrm{max}}$)
  else:
    $\mathbf{p}$ = partial_rand_p($E_{k}^{\mathrm{max}}$)
  
  ! NVE MD for $M$ steps with time step $dt$
  ! Starts from $(\mathbf{s},\mathbf{p})$, ends at $(\mathbf{s}',\mathbf{p}')$
  $(\mathbf{s}',\mathbf{p}')$ = MD_traj($\mathbf{s}$, $\mathbf{p}$, $V$, $\mathbf{h}_0$, $M$, $dt$)
  
  $\mathbf{p}'\leftarrow -\mathbf{p}'$           ! reverse momenta: ensures 
             ! detailed balance
  
  if ($H\left( \mathbf{s'},V,\mathbf{h}_0 ; \mathbf{p'} \right)<H^{\mathrm{sup}} $ AND $E_k(\mathbf{p'})<E_k^0$):
    $(\mathbf{s},\mathbf{p}) \gets (\mathbf{s}',\mathbf{p}')$ ! Accept proposal $(\mathbf{s}',\mathbf{p}')$
  
  $\mathbf{p}\leftarrow -\mathbf{p}$           ! reverse momenta
  
  return $(\mathbf{s},\mathbf{p})$ 
end subroutine
\end{alg}

\begin{alg}[caption={[asdasd] Full momentum randomization from the uniform distribution inside the hyper-sphere $E_k(\mathbf{p})<E_{k}^{\mathrm{max}}$. We set all particle masses to be equal, as described in Sec.~\ref{sec:TE-HMC}. $\mathrm{uranf()}$ is a random number uniform in [0,1]. $\mathrm{granf()}$ is a Gaussian distributed random number, $\mathrm{Normal}(0,1)$.},label=alg:ns_tot_p_rand]
subroutine complete_rand_p$(E_k^\mathrm{max}, m)$
  ! Random vector $\mathbf{p}$ with $E_k(\mathbf{p})<E_{k}^{\mathrm{max}}$.

  ! Generate random unit vector $\mathbf{\hat p}$
  $\mathbf{\hat p}_i = \mathrm{granf()}$ : i = 1, 3N
  $\mathbf{\hat p} = \mathbf{\hat p}/|\mathbf{\hat p}|$
  
  ! Choose random $|\mathbf{ p}|$ with $E_k(\mathbf{p})<E_{k}^{\mathrm{max}}$
  a = $\mathrm{uranf()}$**(1/(3N))
  $\mathbf{p} = a\left( 2 m E_{k}^{\mathrm{max}}  \right)^{\frac{1}{2}}  \mathbf{\hat p}$
  
  return $\mathbf{p}$
end subroutine
\end{alg}

\begin{alg}[caption={[sdfsfd] Partial momentum randomization for TE-HMC nested sampling. Converges to the uniform distribution inside the hyper-sphere $E_k(\mathbf{p})<E_{k}^{\mathrm{max}}$. We set all particle masses to be equal, as described in Sec.~\ref{sec:TE-HMC}. $\mathrm{uranf(}a,b\mathrm{)}$ is a random number uniform in [a,b]. Note, for odd numbers of atoms $N$,  we do not rotate $\mathbf{p}$ component rand\_indices$(3N)$. However,  since rand\_indices($ 3N $) is chosen at random, the subroutine partial\_rand\_p satisfies detailed balance.},label=alg:ns_partial_p_rand]
subroutine partial_rand_p$(\mathbf{p})$
  ! Partial randomization of momentum $\mathbf{p}$.

  ! Choose random $|\mathbf{ p}|$ with $E_k(\mathbf{p})<E_{k}^{\mathrm{max}}$
  $ a = \mathrm{uranf()}$**(1/(3N) )
  $\mathbf{p} = a \left( 2 m E_{k}^{\mathrm{max}}  \right)^{\frac{1}{2}} \mathbf{p}/|\mathbf{p}|$
 
  ! indices $\{1, 2, \dots, 3N\}$ in a random order
  rand_indices = random_order(1,3N) 

  do i = 1, floor(3N/2) ! loop over pairs
    ! pick random angle from $[-\gamma,\gamma]$
    $ \theta = \mathrm{urand(-\gamma,\gamma)}$

    ! pair of components of $\mathbf{p}$ vector
    $ j = \mathrm{rand\_indices}(2i-1) $
    $ k = \mathrm{rand\_indices}(2,i) $

    ! 2D rotation of $\mathbf{p}$ components $j$, $k$
    $ u = \cos\theta\ \mathbf{p}_{j} + \sin\theta\ \mathbf{p}_{k} $
    $ v = -\sin\theta\ \mathbf{p}_{j} + \cos\theta \mathbf{p}_{k} $
    $\mathbf{p}_{j} = u$
    $\mathbf{p}_{k} = v$
  end do

  return $\mathbf{p}$
end subroutine
\end{alg}

\section{TE-HMC Monte Carlo step
\label{app:te_hmc}}

The TE-HMC Monte Carlo step is given in Algorithm~\ref{alg:ns_hmc}.
This algorithm makes use of subroutines for complete and partial momentum randomization given in Algorithms~\ref{alg:ns_tot_p_rand} and~\ref{alg:ns_partial_p_rand} respectively. 

We set all particle masses to be equal, as described in 
Sec.~\ref{sec:TE-HMC}, which helps ensure the sampler spends an approximately equal amount of computer time exploring each degree of freedom.

\section{Potential energy functions}\label{app:pots}

\subsection{Lennard-Jones potential \label{app:LJpot}}

The Lennard-Jones potential used in the paper is the ``truncated and shifted'' potential, given by

\begin{equation} \label{eq:LJpot}
\begin{split}
U\left(\mathbf{r}\right) =& 
\begin{cases} 
4\epsilon \sum_{i<j} \left[ \left(\frac{\sigma}{r_{ij}} \right)^{12} - \left(\frac{\sigma}{r_{ij}   } \right)^{6} - c   \right]  & \: : \: r_{ij}< r_c \\
0 & \: : \: r_{ij} \ge r_c \\
\end{cases}
\\ c =& \left(\frac{\sigma}{r_{c}} \right)^{12} - \left(\frac{\sigma}{r_{c}  } \right)^{6} .
\end{split}
\end{equation}
Here, $r_{ij}$ is the radial distance between particles $i$ and $j$, while two atoms at dynamical equilibrium have a combined energy $-\epsilon$ and are separated by a distance $2^{\frac{1}{6}}\sigma$. 
This potential goes continuously to zero at a radius $r_c$.
Calculations were performed using $\epsilon=1$, $\sigma=1$, $r_c = 3$.

\subsection{Binary Lennard-Jones alloy potential \label{app:BLJpot} } 

The potential used to simulate the binary Lennard Jones alloy is given by

\begin{equation} \label{eq:BLJpot}
U\left(\mathbf{r}\right) =
\begin{cases} 
4 \sum_{i<j} \epsilon_{ij} \left[ \left(\frac{\sigma}{r_{ij}} \right)^{12} - \left(\frac{\sigma}{r_{ij}   } \right)^{6}   \right]  & \: : \: r_{ij}< r_c \\
0 & \: : \: r_{ij} \ge r_c 
\end{cases}
\end{equation}
Calculations were performed using $\epsilon_{AA}=1$, $\epsilon_{AA}=1$, $\epsilon_{BB}=1.5$, $\sigma=1$, $r_c = 3$.
In this potential, all atoms have equal atomic radii, but interactions between different atomic species (A and B) are $1.5$ times stronger than  A--A interactions or B--B interactions.

\subsection{CuAu EAM \label{app:cuaueam}}

We used a Finnis-Sinclair type embedded atom model (EAM) for the
Cu$_x$Au$_{1-x}$ binary alloy.  The potential parameters are from
the method in Ref.~\cite{CuAu_FS_arxiv}, based on the pure element
parameters of Ref.~\cite{Cu_Au_orig_FS}, with inter-species parameters
that are optimized to fit the formation energies of a few crystal
structures at selected compositions of the binary alloy.  The full
parameter set for LAMMPS~\cite{LAMMPS} is available for download from
Ref.~\cite{CuAu_FS_web}.

\subsection{Bead-Spring Polymer Models \label{app:beadspringpot}}

The bead-spring models used to study long molecule chains are based on those
used by Nguyen {\em et al.} to study polymer crystallization~\cite{Karayiannis.2015}.
The energy of a bond between two monomers along the backbone of a polymer chain
\begin{equation} U_b (\ell) = \frac{k_b}{2}(\ell-a)^2 \end{equation} is
harmonic in the distance from the characteristic distance $a.$  The bond
stiffness $k_b=600\epsilon/\sigma^{2}.$  The energy of an angle $\theta$ formed
by three consecutive monomers along the polymer backbone is given by a cosine
potential, \begin{equation} U_a (\theta) = k_a(1-\cos(\theta)), \end{equation}
where the angular stiffness $k_a$ penalizes angular deviations away from a
straight backbone $\theta=180^o$ and is set either to $k_a = 0$ for a
fully-flexible chain or to $k_a = 10\epsilon.$  The nonbonded interaction
between two monomers a distance $r$ from one another is given by Eq.~\eqref{eq:LJpot}
The monomer diameter $\sigma=2^{-1/6}a,$ so that the
bond distance and diameter are commensurate.  The cutoff is set to $r_c =
3\sigma.$  As the name suggests, nonbonded interactions apply only to monomers
that are not bonded together.

\section{Minimum cell depth $d_0$ \label{app:settingd0}}

\begin{figure}[h!]
\centering
\includegraphics[width=1.0\columnwidth]{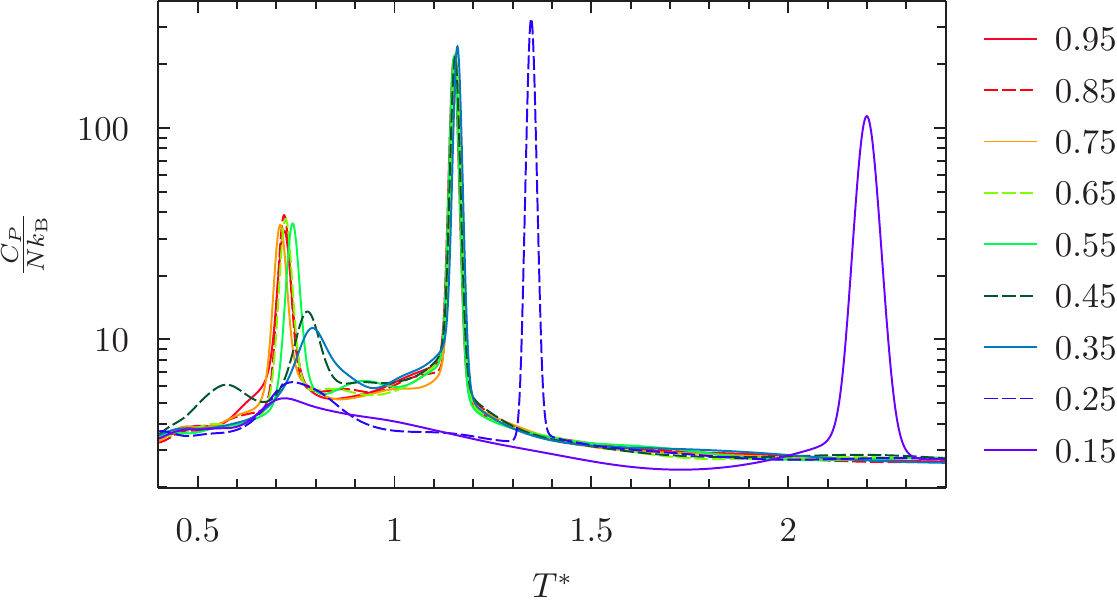}
\caption[Convergence of heat capacity with respect to minimum cell depth, $d_0$.]{
Convergence of the heat capacity $C_p(T)$ with respect to minimum cell depth, $d_0$, for a periodic system of 64 Lennard-Jones particles at pressure $\log_{10}{P^*}=-1.194$.
The peak at high temperature corresponds to condensation, while the peak at lower temperature corresponds to freezing.
The legend on the right shows the value of $d_0$ used in each calculation.
}
\label{fig:mch_cp_converge}
\end{figure}

Figure~\ref{fig:mch_cp_converge} shows the heat capacity of a periodic system of 64 Lennard-Jones particles at fixed pressure.
This calculation used a potential similar to that given in Appendix~\ref{app:LJpot}.
In particular, we set the radial cutoff $r_c=3\sigma$, and $c=0$ in Eq.~\eqref{eq:LJpot}.  We also incorporated the standard long range correction to the energies to account for interactions beyond the cutoff~\cite{FrenkelBookLongrangeEnergyCorrection}.
Each curve corresponds to a single NS simulation, performed using SP-MC nested sampling, with $K=640$, $K_r=1$, $L=2824$, and MC steps in the ratio (1 64-particle SP-MC sweep : 10 cell volume : 1 cell shear : 1 cell stretch). In these calculations each 64-particle SP-MC sweep was broken up into 64 individual single atom SP-MC moves, interspersed between cell moves.
In each calculation we constrained the cell depth to be greater than some minimum value, $d_0$ (see Eq.~\eqref{eq:fixedP_ns_dist}).
A clear transition to a quasi-2D system is observed when reducing $d_0$.
The location of the condensation transition is independent of $d_0$ for $d_0 \geq 0.35$, and the location of the freezing transition for $d_0 \geq 0.65$.

At low values of $d_0$ the simulation cell becomes very thin in at least one dimension and the system's behavior is dominated by unphysical correlations introduced by the periodic boundary conditions. 
The effect of unphysical correlations is reduced at lower densities, and also by increasing $d_0$ which constrains the simulation cell to more cube-like cell shapes.
A larger value of $d_0$ is thus required at higher densities to sufficiently reduce the unphysical correlations.
At the same time, setting $d_0$ too close to 1 excludes crystal structures that require a non-cubic simulation cell.
Quoting from~\cite{pt_phase_dias_ns}, `The window of
independence from $d_0$ grows wider as the number of particles is
increased. For larger numbers of atoms, there are more ways to
arrange those atoms into a given crystal structure, including in
simulation cells that are closer to a cube. Similarly, unphysical
correlations are introduced when the {\em absolute} number of
atoms between faces of the cell becomes too small, and
therefore larger simulations can tolerate ``thinner'' simulation
cells $\mathbf{h}_0$.'
It is clear  from Figure~\ref{fig:mch_cp_converge}  that by imposing a suitable minimum cell height we can remove the unphysical behavior from the fully flexible cell formulation.

\end{document}